\documentclass[10pt,english,journal]{IEEEtran}
\usepackage[utf8]{inputenc}
\usepackage[table,xcdraw]{xcolor}
\usepackage{multirow}
\usepackage{multicol}
\usepackage{graphics}
\usepackage{hyperref}
\usepackage{booktabs}
\usepackage{rotating}
\usepackage{amsmath}
\usepackage{comment}
\usepackage[per-mode=symbol]{siunitx}

\usepackage{float}
\usepackage{caption}
\usepackage{enumitem}
\usepackage{listings}
\usepackage{amsmath}
\usepackage{subfigure}
\usepackage{epsfig} 

\usepackage{array}

\usepackage{url}
\usepackage{graphicx}

\begin{document}

\title{Toward 6G-enabled Brain Computer Interfaces: Technical Requirements, Use Cases, Challenges, and Future Trends}

\author{Houda Hafi, Bouziane Brik,~\IEEEmembership{Senior~Member,~IEEE}, Nuraini Jamil, Abdelkader Nasreddine Belkacem, ~\IEEEmembership{Senior~Member,~IEEE}.
\thanks{ Houda Hafi is with the Faculty of New Information and Communication Technologies, Abdelhamid Mehri University, Constantine, Algeria, e-mail: (houda.hafi@univ-constantine2.dz).}
\thanks{Bouziane Brik is with the Computer Science Department, College of Computing and Informatics, Sharjah University, Sharjah, UAE, e-mail: (bbrik@sharjah.ac.ae).}
\thanks{ Nuraini Jamil is with the Faculty of Computing and Informatics, University of Malaysia Sabah, Malaysia, e-mail: (ainjamil@ums.edu.my).}
\thanks{Corresponding author: Abdelkader Nasreddine Belkacem is with the Department of Computer and Network Engineering, College of Information Technology, United Arab Emirates University, Al Ain, UAE, e-mail: (belkacem@uaeu.ac.ae).}
}

\maketitle

\begin{abstract}
Brain computer interface (BCI) enables the brain to directly control an external device by converting neural signals into actionable outputs. However, effective real-time translation of brain activity strongly depends on the quality of neural communication between the brain and the external device. 6G is the next generation of wireless communication, expected to provide unprecedented levels of data rates, data security, and automation capabilities. In this context, integrating 6G into BCI systems would not only enhance the performance of brain-device communication, but would also create new opportunities for innovative applications. This work provides a comprehensive study on how BCI technology can be built effectively on top of 6G wireless networks by introducing several technical aspects and use cases. We first provide an overview of BCI and 6G, following their progression from early development to convergence through cognitive communication and advanced neural interfaces. We then highlight the need for the upcoming 6G systems toward BCI technology in every aspect, including 6G technologies such as intelligent edge and zero-touch networks, and 6G use cases such as digital twin, immersive communication, and internet of minds. Furthermore, we identify key technical challenges, open issues, and future research directions related to the 6G-enabled BCI paradigm.
\end{abstract}


\section{Introduction}
\label{sec:Introduction}

\subsection{Context and Motivations}
\IEEEPARstart{B}{rain}-computer interface (BCI) or brain machine interface (BMI) technology has come a long way in recent years, enabling direct communication between computers and other devices~\cite{ref1,survey1}. BCI servers to obtain and process neural information in real time by converting neural signals into commands. BCIs can be used in various application areas, especially in medicine, to facilitate control and communication for healthy individuals and those with motor disabilities or cognitive impairments. These BCI applications include adding prosthetics, robotic arms, wheelchairs, augmented reality (AR), and virtual reality (VR) systems. Likewise, BCIs are being used in gaming, learning, and in diagnostic and neuromonitoring settings, such as mental health assessments and sleep studies through real-time monitoring of brain activity~\cite{ref3}.\\
BCI technology offers considerable promise in enhancing the well-being of older people, adults, children with autism spectrum disorder (ASD) and attention deficit/hyperactivity disorder (ADHD), as well as people with neurodegenerative and developmental disorders~\cite{belkacem2020brain, belkacem2023closed, kinney2023pediatric}. Recent improvements in BCI technology have made it possible for non-invasive systems to send data at speeds of over 6–12 bits per second. With these improvements, real time control of external devices such as prosthetics, computer cursors, or robotic systems is becoming possible, with response times that approximate the natural human response ~\cite{ref4}. In addition, invasive BCIs have shown even higher data rates, possibly going over 100 bits per second in experimental settings. This means that communication between the brain and external devices can occur faster and more accurately ~\cite{ref5}. This progress shows that BCIs have the potential to change the way people and computers interact by making communication more direct. In light of these developments, there is a growing trend toward integrating BCI technology into communication networks.

At present, both academia and industry are starting research projects to help shape the future of communication systems, especially 6G. This means going beyond what 5G can do by focusing on the main technological pillars that make ultra-fast, smart, and immersive connectivity possible ~\cite{ref6,ref7}. Despite the lack of clear and well-established key applications and technologies for 6G, there is a strong expectation that 6G will provide hyper-connectivity and hyper-coverage, supporting the development and deployment of new services with stricter requirements, such as fully immersive extended reality (XR), high-fidelity mobile holograms, and digital twins~\cite{ref6}. These features are expected to greatly improve the experiences of users and, even more ambitiously, reshape habitual patterns of human life.

BCIs face several challenges with the previous wireless systems such as ultra-low latency, high bandwidth for data transmission, massive connectivity, and reliable and secure transmission. For example, to enable applications such as neuroprosthetics, brain-controlled drones, or real-time neurofeedback, BCIs must maintain real-time interaction between the brain and external devices, such as exoskeletons, prosthetics, or virtual environments. Furthermore, for accurate control and decoding of brain activity, BCIs require huge bandwidth to efficiently process and transmit the large volumes of neural data produced by BCI devices. This is essential for applications like high-resolution brain monitoring and controlling complex systems. Moreover, BCIs require the ability to connect numerous sensors and devices simultaneously, especially when integrated into large-scale Internet of Things (IoT) systems such as smart homes, healthcare facilities, or brain networks. Finally, BCIs deal with highly sensitive neural data that need to be transmitted securely to prevent privacy breaches or unauthorized access. Therefore, integrating 6G with BCI systems is necessary to unlock the full potential of BCIs. The distinctive capabilities of the future 6G networks, including sub-millisecond latency, terahertz spectrum, massive machine-type communications (mMTC) and advanced security protocols, such as quantum cryptography, will enable the realization of all these applications~\cite{ref6, ref7, ref8}.

\subsection{Review of Existing Surveys}
Several studies have already addressed BCI technology, highlighting its multiple applications and investigating potential benefits and drawbacks. For example, in~\cite{survey1}, a systematic review that examines BCI technologies by comparing them with wireless communication systems is proposed. The key challenges in BCI, such as brain channel modeling, modulation, signal processing, and detection, are highlighted. The review covers both recent and foundational research, with a focus on typical BCI applications, including the Quality-of-Experience (QoE) metric and the concept of the Internet-of-Brains (IoB). The goal of this review is to provide a comprehensive overview of BCI-driven communication on two fronts: BCI as a communication system in itself, and the future integration of BCI with wireless communication as a unified system.

A book chapter provides a thorough exploration of wireless BCI systems and 6G technology in~\cite{survey2}. It emphasizes the application of AI-driven approaches to tackle the security and privacy challenges that emerge with the deployment of 6G networks in the context of wireless BCIs. Specifically, this chapter gives an overview of 6G and WBCI technologies and shows AI-powered schemes' benefits to address privacy and security issues caused by the 6G network deployment on top of BCI. Another survey study is proposed in~\cite{Survey3} to highlight the growing accessibility of BCI devices, driven by advancements such as passive electrodes, wireless headsets, adaptive software and reduced costs. The paper surveys various invasive and non-invasive brain signal acquisition methods, including electrocorticography (ECoG), electroencephalography (EEG), magnetoencephalography (MEG), and magnetic resonance imaging (MRI). In addition, it reviews machine learning and pattern recognition techniques used to interpret brain signals for application control. The work provides a detailed comparative analysis of current BCI techniques, discusses feature extraction and classification algorithms, and offers potential future directions for the field.

In~\cite{ref2}, the authors studied the trend of BCI technology to show the main directions of BCI and where practitioners and research should invest their efforts. The authors have analyzed more than 25,336 metadata of BCI papers from Scopus to study the global trend of BCI field. They showed that the BCI trend is growing exponentially in China since 2019 onward, the United States. Therefore, the authors discussed the main reasons and implications of this trend. Moreover, the main challenges and threats related to BCI technology are also described. The authors designed a typical BCI architecture to address two main BCI threats, security and privacy, as to make the technology commercially reliable to the society.

Having focused on people with severe disabilities, a book chapter was proposed in~\cite{survey33}. This chapter reviews recent studies related to BCI, along with their most relevant applications related to BCI systems. The authors also present the open issues and challenges of the BCI systems and then provide some future directions.

To assess the acceptance rate of people with locked in syndrome about the use of neural signals to communicate (communication BCI), the authors conducted a survey study on people's preferences about the applications they would like to control with a BCI in~\cite{survey4}. The authors interviewed 28 individuals, with locked-in syndrome, during a 3-hour home visit using multiple-choice questions. People with sudden onset disorders preferred to be informed about BCIs and assistive technology at the moment they will need it. Thus, this study provides valuable information to stakeholders in BCI and other assistive technology development.

To gather research activities initiated by both academia and industry to study the integration of 6G with BCI technology, state-of-the-art provided several technical works describing new services that can be supported when using 6G networks to improve the user experiences in~\cite{survey7}. For instance, in BCI+VR Rehabilitation Design, one work has designed a closed-loop virtual-reality (VR), motor imagery (MI) rehabilitation training system based on BCI, aiming to enhance cognition, self-control, and emotional regulation of drug addicts through personalized rehabilitation schemes.

\begin{table*}[]
\caption{Existing survey studies on BCI. \textbf{H: High, M: Medium, and L: Low.}}
\label{T1}
\begin{tabular}{|c|c|c|c|c|l|}
\hline
\cellcolor[HTML]{EFEFEF}\textbf{Works}                  & \cellcolor[HTML]{EFEFEF}\textbf{BCI Survey}                & \cellcolor[HTML]{EFEFEF}\textbf{6G-Enabled BCI}         & \cellcolor[HTML]{EFEFEF}\textbf{\rotatebox{90}{6G-enabled Projects for BCIs}} & \cellcolor[HTML]{EFEFEF}\textbf{\rotatebox{90}{Future Directions}} & \multicolumn{1}{c|}{\cellcolor[HTML]{EFEFEF}\textbf{Contribution}}                                                                                                                                                                                                                                                   \\ \hline
\cite{Survey3} & \cellcolor[HTML]{F8A102}H          & \cellcolor[HTML]{96FFFB}L          & \cellcolor[HTML]{96FFFB}L                          & \cellcolor[HTML]{9AFF99}M                      & \begin{tabular}[c]{@{}l@{}}A survey paper on various invasive and non-invasive brain \\ signal acquisition methods such as, ECoG, EEG, MEG, and MRI.\end{tabular}                                                                                                                            \\ \hline
\cite{ref2}    & \cellcolor[HTML]{F8A102}H          & \cellcolor[HTML]{96FFFB}L          & \cellcolor[HTML]{96FFFB}L                          & \cellcolor[HTML]{9AFF99}M                      & \begin{tabular}[c]{@{}l@{}}A review paper on the trend of BCI technology. It analyzed more than 25336\\  metadata of BCI papers from scopus to study the global trend of BCI field.\end{tabular}                                                                                             \\ \hline
\cite{survey33}    & \cellcolor[HTML]{F8A102}H          & \cellcolor[HTML]{96FFFB}L          & \cellcolor[HTML]{96FFFB}L                          & \cellcolor[HTML]{F8A102}H                      & \begin{tabular}[c]{@{}l@{}}A book chapter on recent BCI studies along with their relevant \\ applications related to BCI systems\end{tabular}                                                                                                                                                \\ \hline
\cite{survey7}    & \cellcolor[HTML]{F8A102}H          & \cellcolor[HTML]{96FFFB}L          & \cellcolor[HTML]{96FFFB}L                          & \cellcolor[HTML]{9AFF99}M                      & \begin{tabular}[c]{@{}l@{}}A guest editorial was initiated to gather works from both industry and academia \\ integrating communication in order to improve on the user experience.\end{tabular}                                                                                             \\ \hline
\cite{survey4}    & \cellcolor[HTML]{F8A102}H          & \cellcolor[HTML]{96FFFB}L          & \cellcolor[HTML]{96FFFB}L                          & \cellcolor[HTML]{96FFFB}L                      & \begin{tabular}[c]{@{}l@{}}A survey study was performed about people preferences \\ on the applications they would like to control with a BCI\end{tabular}                                                                                                                                   \\ \hline
\cite{survey2} & \cellcolor[HTML]{F8A102}H          & \cellcolor[HTML]{96FFFB}L          & \cellcolor[HTML]{96FFFB}L                          & \cellcolor[HTML]{9AFF99}M                      & \begin{tabular}[c]{@{}l@{}}A book chapter provides a thorough exploration of wireless \\ BCI systems and the application of AI-driven approaches\end{tabular}                                                                                                                                \\ \hline
\cite{survey1} & \cellcolor[HTML]{F8A102}H          & \cellcolor[HTML]{9AFF99}M          & \cellcolor[HTML]{9AFF99}M                          & \cellcolor[HTML]{F8A102}H                      & \begin{tabular}[c]{@{}l@{}}A survey paper that examines BCI technologies and comparing \\ them with wireless communication systems.\end{tabular}                                                                                                                                             \\ \hline
\textbf{This survey}            & \cellcolor[HTML]{F8A102}\textbf{H} & \cellcolor[HTML]{F8A102}\textbf{H} & \cellcolor[HTML]{F8A102}\textbf{H}                 & \cellcolor[HTML]{F8A102}\textbf{H}             & \textbf{\begin{tabular}[c]{@{}l@{}}A comprehensive survey on how BCI technology can be effectively built on \\ top of 6G wireless networks. In particular, our study focuses on several technical\\  aspects and use cases in which 6G could enhance the capabilities of BCIs.\end{tabular}} \\ \hline
\end{tabular}
\end{table*}

~\tablename~\ref{T1} illustrates the main topics discussed along the existing survey papers, and compares their contributions with respect to our work, in order to provide an easy understanding of the differentiation features with respect to the state-of-the-art.

Despite the presence of several survey papers discussing BCI technology and its applications, there is a lack of comprehensive surveys jointly investigating 5G/6G communications and BCI aspects, able to effectively explore the potential of 6G networks for enhancing the capabilities of BCIs. In addition, although the integration of BCIs with 5G/6G networks has been addressed, e.g., in~\cite{survey1}, such study focused only on the physical layer in terms of modulation, signal processing and detection, and did not explore the potential of the multiple 6G networks' technical aspects and uses-cases in making BCIs applications more efficient, reliable and secure. Therefore, a comprehensive survey of 6G networks and its potential in designing the future BCI is greatly needed to guide practitioners and researchers.  

\subsection{Key Contributions and Insights}
The main contributions of this article are summarized as follows:
\begin{itemize}
    
    \item \textbf{Bridging the gap between BCI and 6G Networks:} 
    In literature, there are myriads of papers examining BCI~\cite{bablani2019survey, mridha2021brain}, and 6G networks~\cite{jiang2021road, shahraki2021comprehensive} in isolation. However, the interplay between BCI and 6G is completely absent from the discourse. Compared to other surveys in the field, this study fills this gap by uniquely analysing the important contributions that the connection between BCI and 6G can lead to.
   
    \item \textbf{Detailed 6G technical aspects and use-cases for BCI:} A deeper dive into the chief 6G technical aspects for BCI (e.g., intelligent physical layer, intelligent edge computing, resource management) and BCI use cases (e.g., brain digital twin and internet of minds) is taken to investigate how 6G can improve the functionalities of all BCI stakeholders. 
    
    \item \textbf{Towards a successful implementation of BCI over 6G networks:} This work describes several standards/frameworks that relate to the convergence of BCI and the next-generation of networks, with the goal of ensuring reliable and interoperable 6G-enabled BCI solutions. Furthermore, it addresses various open implementation challenges along with their potential solutions, highlighting future directions for the application of BCI in 6G systems.
\end{itemize}

\subsection{Paper Outline}
\begin{figure*}[h!]
	\centering
        \includegraphics[width=1.02\textwidth, height=15cm]{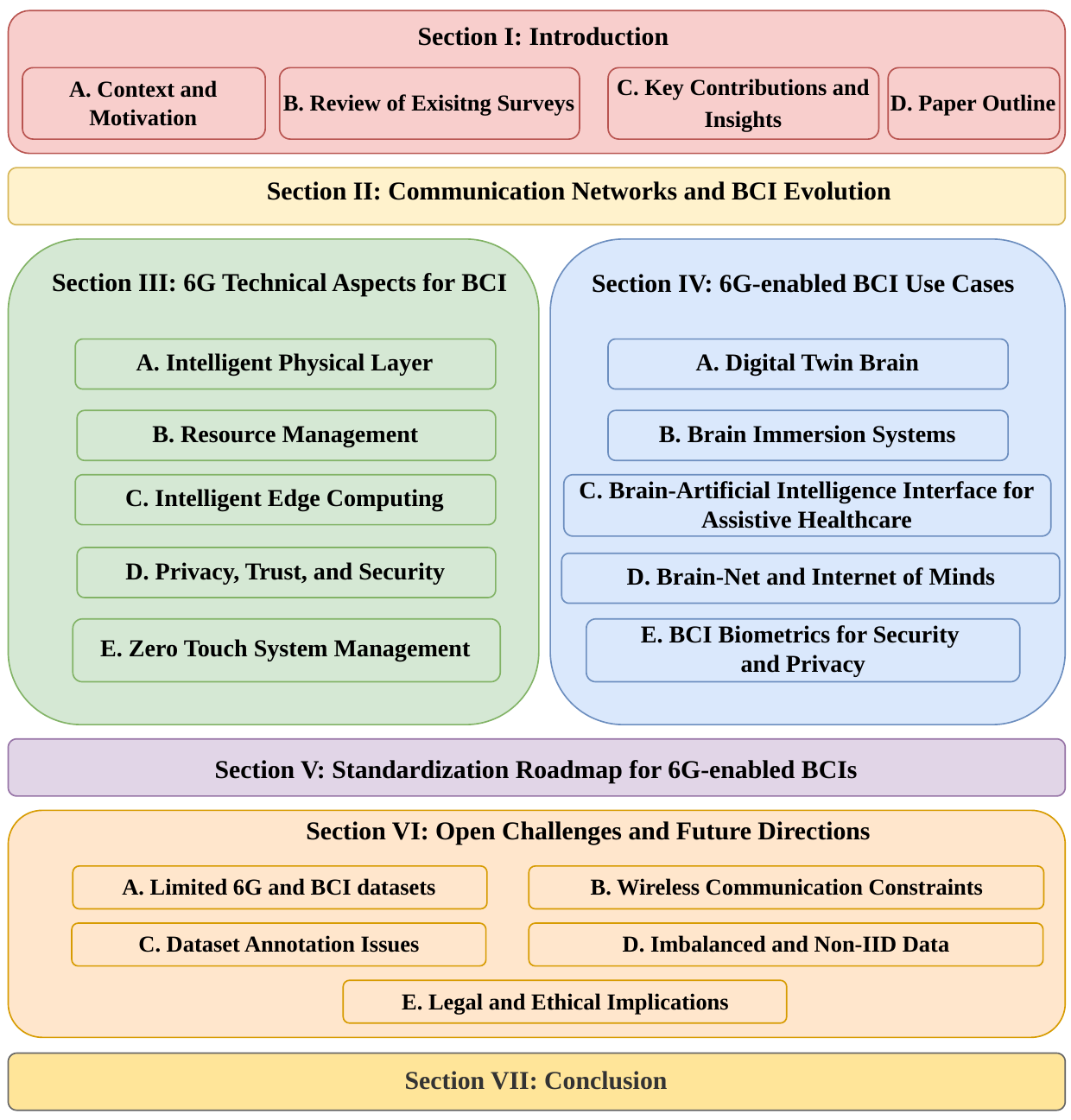}
	\caption{Overview of the contents of this paper.}
	\label{fig:stc_paper}
\end{figure*}

The organization of this paper is described as follows. Section~\ref{sec:Background} discusses the progressive evolution of 6G and BCI technologies. Section~\ref{sec:6G Technical Aspects for BCI} reviews the technical requirements of 6G networks to support the needs of BCI technology. Section~\ref{sec:6G Use Case for BCI} focuses on how to integrate the unique features of the forthcoming 6G networks to deliver the desired functionalities of the BCI use cases and services. Section~\ref{sec:Projects/Standards for 6G-enabled BCI} discusses the 6G-enabled projects/standards. In Section~\ref{sec:Open Challenges and Future Directions}, we distill some future research directions, and finally we conclude the article in Section~\ref{sec:Conclusion}.
To facilitate the reader's navigation through the paper, ~\figurename~\ref{fig:stc_paper} presents the survey roadmap.

\section{Communication Networks and BCI Evolution}
\label{sec:Background}
The origins of BCI research date back to 1875, with the discovery of electrical signals in animals' brains. In 1924, Hans Berger recorded the first electrical signals of the human brain, marking the birth of electroencephalography (EEG). Building on this foundation, the first experimental BCI tests were carried out on monkeys in 1969 and 1970, and the concept was formally introduced in 1973 by Jacques Vidal in the paper titled “Toward Direct Brain-Computer Communications” \cite{vidal1973toward}. At that time, the field of wireless communication was in its earliest analog systems, the zeroth-generation (0G), which were designed exclusively for simple voice transmission and had no connection to BCI technology~\cite{0G}.

In 1980, the first-generation (1G) introduced analog cellular telephony, enabling mobile voice calls over a structured network~\cite{1G}. Around the same period, early wireless EEG experiments were conducted and the first P300 speller was introduced by Farwell and Donchin in 1988~\cite{farwell1988talking}, which demonstrated the potential of EEG-based communication between the brain and external devices.

In the early 1990s, the second-generation (2G) of mobile networks enabled the migration from analog to digital technology. This shift not only improved call quality, but also introduced text messaging (SMS), multimedia messaging (MMS), and picture messaging as additional services~\cite{2G}. During the same period, BCI research also advanced, with the emergence of mobile neuro feedback and simple non-invasive BCI applications. The first attempts with human beings were performed in this period. In 1996, the steady-state visual evoked potential (SSVEP)-based BCI was introduced, and in 1998, Philip Kennedy implanted the first invasive BCI into human.

With the efforts of the International Telecommunication Union (ITU), the third-generation (3G) launched in 2001, optimizes network capacity and supports advanced services, including the Global Positioning System (GPS), mobile television, multimedia communications, and video calls~\cite{3G}. Within this period, significant progress was also achieved in the field of BCI, specifically, the first implantation of a commercial BCI into a human brain in 2004.

Subsequently, in 2009, the fourth-generation (4G) standard was introduced, further enhancing mobile services by enabling Voice over IP (VoIP), online gaming, high-definition (HD) mobile television, mobile web access, and 3D television ~\cite{4G}. In parallel, BCI technology marked important progress toward real-time BCI streaming and neural control. In 2012, researchers successfully developed a method to control a robotic arm by a quadriplegic patient using neural signals ~\cite{grigorescu2012bci}. In 2013, the first demonstrations of brain-to-brain communication were achieved ~\cite{rao2014direct}. Between 2015 and 2016, the concept of Neural Dust emerged ~\cite{neely2018recent}. By 2017, researchers had achieved image reconstruction and speech decoding directly from brain activity ~\cite{takagi2023high, kunz2025inner}.

Currently, several network operators around the world are deploying 5G cellular networks to support more advanced services such as ultra-reliable low-latency communications (uRLLC), enhanced mobile broadband (eMBB) and massive machine-type communications (mMTC). uRLLC ensures response times as low as 1 millisecond. eMBB delivers speeds of up to 10 Gbps. Meanwhile, mMTC facilitates the large-scale deployment of IoT devices, supporting more than 100 times the number of connected devices per unit area compared to 4G. With 5G, the availability and reliability of the network is expected to reach 99.999\%~\cite{5G}. These capabilities enable advanced applications, including AR, VR, mixed reality (MR), the Internet of Things (IoT), autonomous vehicles, and Industry 4.0~\cite{5G_app1}. When combined with BCI technology, these advances pave the way for immersive and real-time BCI applications, allowing users to interact with virtual environments, control devices, or access complex IoT systems directly through neural signals.

Current research efforts are now oriented toward 6G, which introduces new visions, communication concepts, applications, and enabling technologies. This will support the next generation of BCI systems, integrating cognitive, affective, and real-time wireless neural interactions.  

\begin{figure*}[!t]
\centering
  \includegraphics[height=13cm,width=\linewidth]{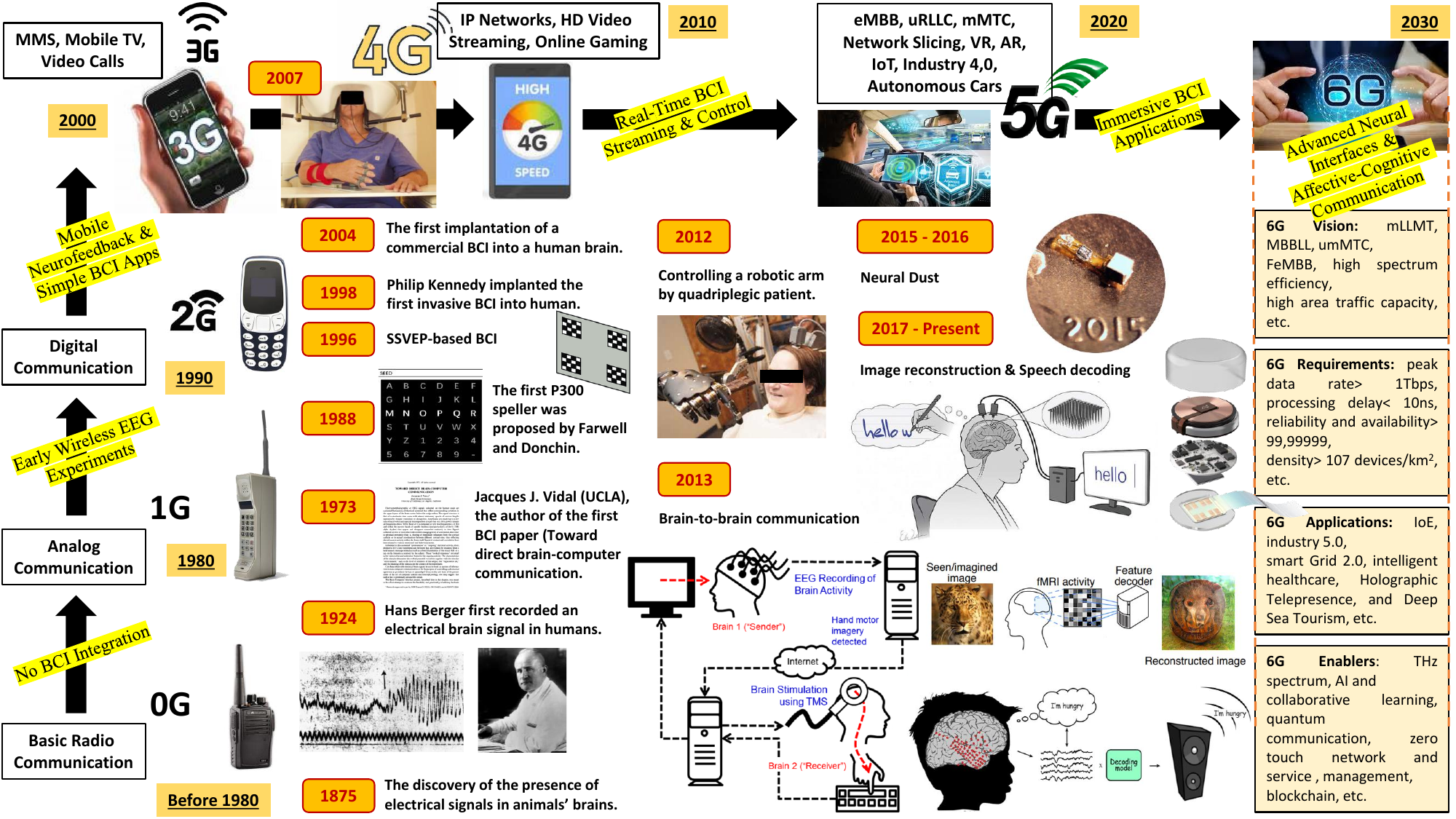}
  \caption{Evolution of mobile networks and BCI technology: From 0G connectivity to 6G cognitive communication.}
   \label{Evo}
\end{figure*}

~\figurename~\ref{Evo} outlines the parallel evolution of cellular networks and BCI technologies, highlighting the progression from 0G without any form of BCI to the incorporation of cognitive communication capabilities and advanced neural interfaces anticipated in 6G.

\section{6G Technical Aspects for BCI} 
\label{sec:6G Technical Aspects for BCI}
In this section, we will explore the technical requirements of 6G networks to support BCI technology, as well as potential solutions to meet these needs.
\subsection{Intelligent Physical Layer}
\subsubsection{Introduction, Requirements and Potential Solutions}
The physical layer is a foundational component of BCI systems, responsible for transmitting and receiving brain signals over a communication channel. Its design directly influences the reliability and performance of BCI applications. Physical layer designers should consider several aspects to meet the diverse performance requirements of BCI use cases.
One of the key requirements would be a high data rate to enable real-time communication between the brain and the computer. Another important factor would be low latency, to minimize the delay between the brain's signals and the computer's response.  Additionally, the physical layer would need to be able to handle a large number of channels, since BCI systems typically involve recording signals from multiple electrodes attached to the scalp or implanted in the brain. The physical layer would also need to provide high signal quality and reliability, to ensure that the recorded signals accurately reflect the activity of the brain. A potential approach to designing the physical layer for 6G BCI technology would be to use millimeter-wave (mmWave) frequencies~\cite{tripathi2021millimeter}. These high-frequency waves can provide the high data rates and low latency needed for BCI applications while also allowing for a large number of channels. However, mmWave frequencies have limited range and are susceptible to signal blockage, so careful antenna design and placement would be important to ensure reliable communication.  Another potential approach would be to use terahertz frequencies (THz), which offer data rates even higher than the mmWave frequencies~\cite{rappaport2019wireless}. THz frequencies also have the advantage of being able to penetrate some types of biological tissue, which could potentially enable noninvasive BCI systems. However, THz frequencies are even more susceptible to signal attenuation and blockage than mmWave frequencies~\cite{basharat2022exploring, yi2022ray}, so significant research would be needed to develop reliable THz-based BCI systems.  Overall, the physical layer of 6G technology for BCI applications would need to balance the competing requirements of high data rates, low latency, channel capacity, signal quality, and reliability. MmWave and THz frequencies are two potential approaches that could be used to achieve these goals, but further research and development will be needed to determine the most effective physical layer design for 6G BCI systems.

The integration of Artificial Intelligence (AI) techniques such as machine and deep learning into the design of 6G physical layer would intelligently enhance the performance of BCI applications. More specifically, AI-enabled 6G physical layer optimizes channel coding schemes, such as error-correcting codes (e.g., LDPC and Turbo codes), based on channel conditions. This enhances BCI data transmission reliability in noisy channels~\cite{sattiraju2018performance}. DL/ML can also optimize modulation schemes adapting them to varying channel conditions in real-time, maximizing BCI data throughput while minimizing error rates~\cite{hall2023deep}. The incorporation of AI algorithms into the 6G physical layer can also assist in advanced BCI signal processing, including noise reduction, interference mitigation and channel estimation~\cite{jagannath2021redefining}~\cite{helal2022signal}. These capabilities can improve the accuracy and reliability of BCI data transmission. Moreover, BCI data is highly sensitive, thus the security of the communication link between brain signals and external devices is paramount. In this context, AI can be used to implement robust encryption and authentication mechanisms at the physical layer to protect BCI data against interception and tampering~\cite{sharma2023deep}. Therefore, by leveraging DL/ML techniques in 6G physical layer, BCIs applications can achieve higher data rates, lower latency, better reliability, and adaptability to dynamic and complex communication environments. 

\subsubsection{Realistic Scenario}
\begin{figure}[!h]
\centering
  \includegraphics[width=\columnwidth, height=5cm]{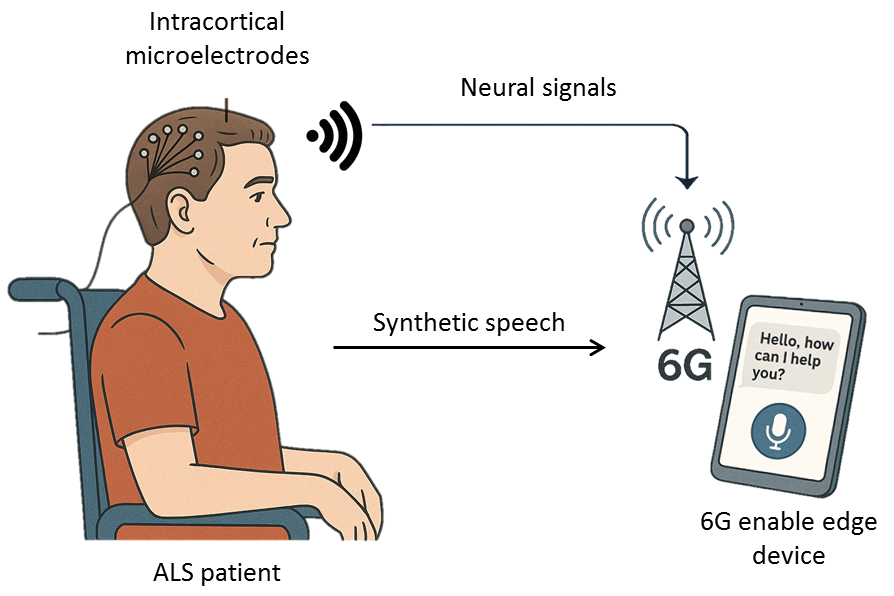}
  \caption{Wireless transmission of neural signals from an ALS patient to a 6G-enabled edge computing device.}
   \label{Scenario1}
\end{figure}

An illustrative example of BCI–6G integration is the assistance provided to patients with Amyotrophic Lateral Sclerosis (ALS) who have lost their ability to speak or move. In this scenario, the patient uses a BCI communication device with intracortical microelectrode implants that record neural signals linked to intended speech. These signals are wirelessly transmitted to a 6G-enabled edge computing device such as a hospital server or smart tablet that decodes the neural activity into text and converts it into synthetic speech, enabling real-time communication (Fig.~\ref{Scenario1}). The intelligent 6G physical layer autonomously selects the optimal frequency band, minimizing interference and maintaining low-latency, high-reliability connections. For instance, if the hospital’s Wi-Fi network becomes congested, the system automatically switches to a less crowded millimeter-wave (mmWave) or terahertz (THz) band to sustain stable transmission.

Machine learning algorithms at the physical layer predict the most suitable frequency spectrum based on time, network load, and environmental factors, preventing congestion and ensuring consistent connectivity. In cases of sudden interference from devices like MRI machines or infusion pumps, the 6G layer instantly redirects transmissions to another band to avoid signal degradation or data loss. Given the limited power of implanted BCIs, the intelligent layer also optimizes energy efficiency, reducing retransmissions and preserving battery life. This adaptive 6G framework enables seamless, real-time communication between ALS patients, caregivers, and clinicians, demonstrating how 6G-enhanced BCIs can deliver reliable, high-speed, and interference-free neural data transmission in healthcare applications.

\subsection{Resource Management}
\subsubsection{Introduction, Requirements and Potential Solutions}
Given the real-time nature of BCIs and the need for reliable data transmission between the brain and external devices, intelligent and adaptive management of 6G bandwidth becomes a key priority. In this sense, AI-driven bandwidth allocation enables real time adaptation to mitigate probable network congestion~\cite{lin2021ai}, while cognitive radio technologies increase the efficient utilization of the spectrum by exploiting the existence of underutilized
radio~\cite{nasser2021spectrum}.  In addition, terahertz communication provides ultra-high data rates suitable for dense, high-resolution neural signal transmission~\cite{song2021terahertz}. Strategies such as interference-aware resource control, intelligent neural data compression and prioritization, as well as edge-assisted brain data processing, further reduce end-to-end latency and alleviate core network congestion~\cite{cheng2023ai}. Implementing a combination of these techniques optimizes bandwidth utilization for various BCI applications requiring high data rates and low latency.

Moreover, 6G requires efficient mobility management for BCI systems. BCI users are not always stationary and may exhibit significant mobility, especially with the use of mmWaves and Terahertz technologies, through the adoption of small Base Stations (BSs), such as Femto,  Pico and drone-based cells which increases the probability of handoff events~\cite{9195500}. Consider a scenario where a wearable EEG device in a 6G network continuously monitors and processes brainwave activity for a mobile user in real-time at the edge. At some point, the user will disconnect from the coverage area of the current cell and need to establish a connection with a new one. Therefore, service migration of the ongoing session from the former edge server to another must be considered in the design of 6G edge-assisted BCI application~\cite{mollel2021survey}.

It is also important to note that each 6G-enabled BCI use case will require customized and, at times, conflicting performance metrics compared to others. To this end, 6G introduces network slicing concept that involves the creation of multiple virtual networks or ``slices" on a single physical infrastructure, with each slice comprising customized resources (memory, computing, and network) and functions (virtual network functions) to support specific services~\cite{khalili2019network}.  
In the context of 6G-enabled BCI, network slicing can generate specialized slices tailored for real-time transmission of brain signals, ensuring high dependability, low latency, and low packet loss. The network infrastructure must be created with flexible and programmable architectures to allow network slicing in 6G networks for BCI. Some examples of such designs are software-defined networking (SDN) and network services virtualization (NFV). When SDN and NFV are combined, network operators can design network slices that are adaptable, adjustable, and programmable for BCI applications \cite{xie2022practically}. The slice can also be tuned to particular constraints of the BCI application, such as low power consumption, fast data throughput, and low packet loss. For example, the slice may take advantage of a millimeter-wave wireless link with beamforming to establish connections between the EEG headset and the processing platform characterized by high bandwidth and low latency. The slice also can prioritize BCI traffic over other traffic on the network, which helps to ensure that the performance of the BCI system is not negatively impacted by the performance of other applications running on the network.  However, managing the allocation of resources to these different slices is a challenging task. In this respect, AI-assisted approaches that encompass the entire slice lifecycle, from admission control~\cite{Ojijo20}, resource orchestration~\cite{Rezazadeh23} to radio scheduling~\cite{Setayesh22} could be applied. In the literature, numerous research works have focused on addressing this challenge~\cite{shen2020ai,bega20}. 

Likewise, managing transmission power is crucial for reliable communication between the user's BCI device and external devices, as low power results in weak signals causing communication errors, while high power causes interference and energy wastage. Therefore, dynamic solutions that adapt to individual BCI users' needs and changing 6G network conditions are essential. In this context, ML/DL such as deep reinforcement learning and optimization techniques can play a vital role in controlling power allocation to ensure good signal quality, minimize interference, and conserve energy~\cite{nasir2019multi}.
In~addition, to process brain signals, BCI applications require a large amount of computational power~\cite{netzer2020real}, apart from the edge computing, the integration of Quantum Computing (QC) into 6G networks would offer several benefits to 6G-enabled BCI systems. QC has the potential to revolutionize both the communication and computing perspectives of 6G network~\cite{wang2022quantum}. For instance, using the next generation of quantum computing processors can significantly enhance the speed and efficiency of 6G network operations, enabling faster BCI data processing. Furthermore, quantum communication can make 6G networks more secure using Quantum Key Distribution (QKD)~\cite{muheidat2022security}. This advancement plays a vital role in ensuring the integrity and confidentiality of brain signals during transmission. QC can also help in optimizing the allocation of 6G wireless resources in the network. For example, if a 6G network needs to allocate specific frequencies to different BCI users, devices, or services to ensure efficient and reliable communication, finding the best allocation is computationally challenging. QC processing capabilities could quickly analyze all possible frequency allocation combinations, leading to the best solution that minimizes interference and optimizes network performance. Even more, the synergy between ML/DL algorithms and QC leads to a new paradigm called Quantum Machine Learning (QML)~\cite{biamonte2017quantum} that can be employed to enhance AI-driven 6G network management, improving AI's decision-making capabilities in 6G network optimization. As well, combining ML and QC enables rapid processing and analysis of large and complex neural datasets, leading to improved precision and accuracy in decoding the BCI user's intentions. 

\subsubsection{Realistic Scenario}
Imagine a person who has lost the ability to move their arms but wants to use a robotic arm to carry out simple daily tasks—like picking up a glass of water or opening a door—just by thinking. To do this, they wear a special headset that can read their brainwaves. Fig.\ref{Scenario2} illustrates these brain signals are sent through a super-fast and reliable internet connection (enabled by future 6G networks) to a powerful computer located in the cloud. This computer quickly understands what the person is trying to do—like reach for something—and sends the right instructions to the robotic arm to carry out that action.

To make sure everything works smoothly, the system uses a reserved part of the internet that is specifically set aside for these brain-control tasks. This helps avoid delays and ensures the robotic arm responds almost instantly. Since brain signals are very personal and sensitive, the system also has strong privacy and security features in place to protect the user data. The cloud computer uses smart technologies to learn the user's behavior over time. For instance, if the person's brain activity becomes more complex—perhaps when they’re trying to perform a new or difficult task—the system can automatically give more brainpower and memory to make sure everything still works perfectly. This setup allows people to control a robotic arm using just their thoughts, giving them more independence and improving their quality of life.

\begin{figure}[!h]
\centering
  \includegraphics[width=\columnwidth, height=5cm]{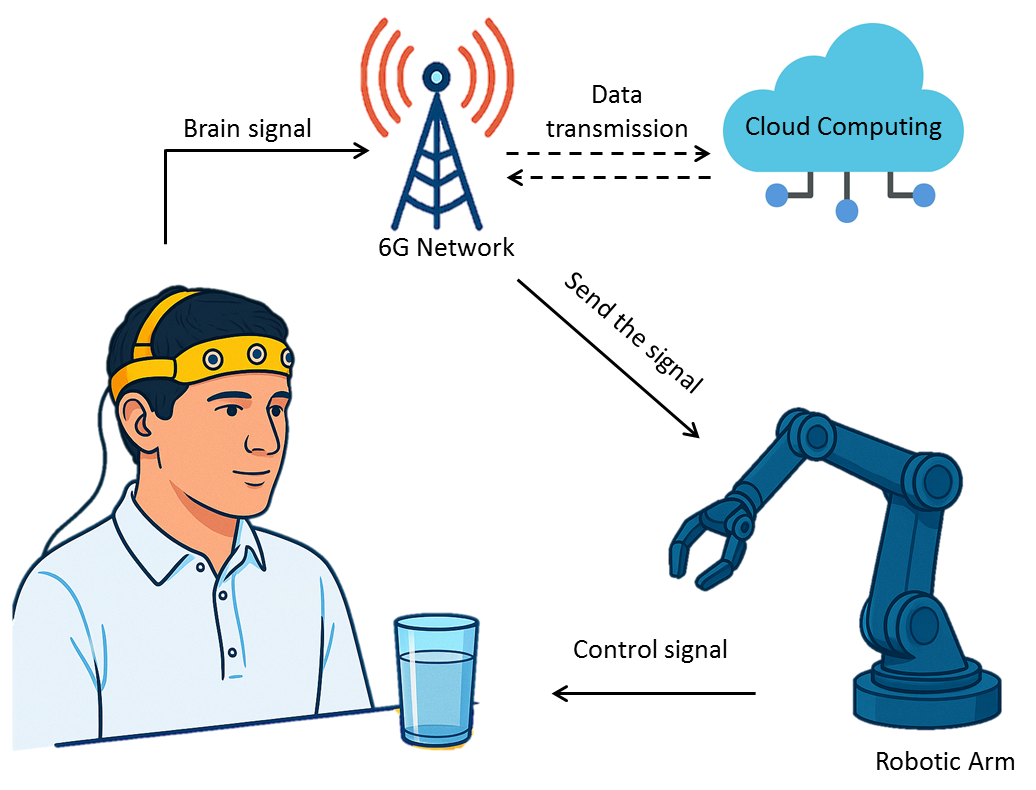}
  \caption{Thought-controlled robotic arm powered by 6G and Cloud Intelligence}
   \label{Scenario2}
\end{figure}

\subsection{Intelligent Edge Computing}
\label{sec:IEC}
\subsubsection{Introduction, Requirements and Potential Solutions}
To gain a deep understanding of brain activity and its correlation with various cognitive and clinical factors, BCI applications generate massive amounts of data (e.g., high-resolution brain imaging), reaching terabytes and even petabytes, requiring specialized infrastructure for storage and analysis. Transmitting this BCI data to cloud servers can lead to several challenges, among others data explosion, bandwidth consumption, intermittent connectivity, and security issues. Moving computing resources nearer to the brain data source, aka Multi-Access Edge Computing (MEC), serves as a viable solution for addressing these challenges~\cite{al2019edge}. Furthermore, intelligent edge computing (IEC), which consists of the integration of AI capabilities at the edge, is expected to be a crucial component in future BCI systems, making them more intelligent, secure, and efficient. IEC allows for real-time processing of the brain signals directly at the edge, rather than being sent to a remote server. This can significantly reduce latency and improve the performance of real-time BCI applications, such as assistive technology. Moreover, this would improve the User eXperience (UX) by providing reactive and responsive interactions with BCI systems, which is critical in gaming, virtual reality, and other applications. BCI-based VR applications may lead to seamless user experience wich means that from the user’s perspective, there is one mode of operating the BCI. In other words, the user does not have to consciously switch between operating one type of BCI versus another.
Nevertheless, it is important to emphasize that BCI devices are designed to read individual's brain activity, which is among the most personal and sensitive data. Therefore, forwarding these neural data to edge-based training can indeed trigger significant data privacy concerns during the transmission process. To minimize the risk of data breaches or unauthorized access, Federated Learning (FL) techniques can be used. In FL settings, each BCI device trains a local model while contributing to a global model~\cite{zhang2021survey}. In other words, each device participates in the training process without sharing its own data. Without a doubt, this will significantly reduce the bandwidth usage and enhance data privacy, however, training the full model on BCI devices can lead to a premature termination of the training process due to their limited resources. In this context, new distributed machine learning methods, namely Split Learning (SL) and Split Federated Learning (SFL), could be applied. Unlike FL, SL and SFL consist of running only a section of the model locally and transmitting smashed data to the edge to continue the training process~\cite{hafi2024split}. This will extend the battery life of BCIs devices. Therefore, the combination of EIC and BCI will play a vital role in enhancing the performance and capabilities of BCI systems in various applications. 
\subsubsection{Realistic Scenario} 

Imagine a person with a spinal cord injury who has lost control of all limbs and participates in a clinical trial using a BCI to operate a robotic exoskeleton. With 6G-enabled edge computing, the user’s brain signals are processed in real time with minimal latency, allowing smooth and responsive control of the exoskeleton. The edge server is positioned near the user, ensuring rapid signal analysis without relying on distant cloud servers. Equipped with high-performance CPUs and GPUs, the edge device efficiently processes brain impulses through personalized machine learning algorithms that translate neural signals into control commands. Adaptive resource allocation further enhances the system by dynamically adjusting computing power according to signal complexity, ensuring precise and reliable exoskeleton operation even during demanding tasks.

\subsection{Privacy, Trust, and Security}
\subsubsection{Introduction, Requirements and Potential Solutions}
BCI devices record highly personal data related to individual thoughts, intentions and emotions. Therefore, unauthorized access (for intentional or accidental reasons) to this information can result in a serious violation of mental privacy, leading to significant physical, psychological, and ethical implications. For example, in a BCI-controlled ventilator scenario, if the communication channel between the BCI and the ventilator is not adequately secured, an attacker with malicious intent could manipulate the BCI system to change the ventilator settings (altering the oxygen flow rate or even turning the ventilator off) leading to life-threatening situations. Therefore, the implementation of robust solutions that protect the data collected by BCI devices, at rest and in transit, is of utmost importance. 

Many emerging and innovative solutions could be used to address these challenges; among others, we cite cryptography-based mechanisms such as Homomorphic Encryption (HE) algorithms that allow computations on encrypted BCI data (e.g., EEG data) without decryption~\cite{liu2020classification}. Applying this technique can ensure that brain signals remain confidential throughout the processing pipeline. Another cryptographic protocol is the Secure Multi-Party Computation (SMPC) that ensures data confidentiality and integrity during collaborative data processing. For example, in~\cite{senanayake2022neurocrypt}, authors introduce a novel learning approach for distributed neuroimaging data based on SMPC called NeuroCrypt. The idea behind the work is to enable organizations to collaborate securely on machine learning model training without sharing their data in plaintext and without relying on a secure aggregator (in a peer-to-peer setting). Results indicate that NeuroCrypt achieves the highest accuracy in fewer or similar iterations and matching the accuracy of the model trained on plaintext data. Although both HE and SMPC enable computations on encrypted BCI data, they are computationally intensive and time-consuming, making real-time and efficient processing of high-dimensional BCI data challenging. Embedding secure processors into BCI devices or cloud/edge servers where data processing is performed is another method that provides a secure execution environment to prevent unauthorized access to BCI data by performing computations on trusted hardware, e.g., Intel Software Guard Extensions (SGX)~\cite{anati2015intel}. However, this may be expensive, complex, and energy intensive. Another viable solution is data perturbation, which consists of obfuscating the original neural data by adding some noise while maintaining data utility~\cite{rahman2023efficient}. In this context, we present two techniques: Differential Privacy (DP) and the process of data reconstruction. The application of the DP technique to brain signals (e.g., EEG signals) enables researchers to study patterns in the EEG data to develop ML/DL models (e.g., epileptic seizure prediction or emotion recognition) while withholding exact details of each individual EEG signals in the dataset. Although this technique offers robust privacy and upholds individuals' rights, it should be applied with careful consideration because adding excessive noise can reduce data utility. Therefore, it is important to strike a balance between data utility and privacy preservation.  Beyond DP, data reconstruction is another approach in the data perturbation category. It typically uses two strategies, namely Dimensionality Reduction (DR) and synthetic data generation. DR approach consists in reducing the number of features (dimensions) in the BCI data while retaining as much relevant information as possible~\cite{cozza2020dimension}. In other words, the objective is to find a smaller-dimensional representation of the BCI data that captures its essential structure. However, choosing the right dimensionality reduction technique (linear, nonlinear, statistical, etc.) for a BCI application can be challenging, especially for large-scale data~\cite{reddy2020analysis}. Likewise, certain dimensionality reduction methods may not ensure adequate privacy protection, attackers could still infer or reconstruct sensitive information, even after reducing data dimensionality. Therefore combining more advanced privacy-preserving techniques like DP or SMPC should be considered during dimensionality reduction. 

Synthetic data generation is another commonly adopted strategy that involves the creation of artificial BCI data to mimic the features of the original data while removing or obfuscating sensitive details such as personal identifiers. Among the methods that can be used to generate synthetic BCI data, we cite Generative Adversarial Networks (GANs)~\cite{debie2020privacy}. For example, in~\cite{pascual2020epilepsygan} authors proposed EpilepsyGAN to generate high-quality synthetic seizure-like brain electrical activities that can be used to train seizure detection algorithms, eliminating the need for sensitive recorded data. This technique allows researchers to work with synthetic data while securing the original BCI data and protecting individuals' privacy. In addition to the strategies we have already discussed, it is essential to consider privacy-preserving ML/DL techniques. Conventional BCI model training relies on the transmission of BCI data to a central server~\cite{zhang2019survey}. Apart from concerns related to bandwidth consumption and the single point of failure, this could lead to significant security and privacy issues. Federated learning is a distributed and collaborative approach that enables local training of BCI models without compromising data privacy~\cite{ghader2023exploiting}. However, federated learning is not resistant to all privacy attacks, especially model-targeted attacks, as both the clients and the server have full access to the local and global models. Therefore, alternative distributed training methods that prioritize model privacy more effectively, such as split learning and split federated learning, are preferred. SL and SFL algorithms divide the ML/DL-enabled BCI model into multiple smaller sections and train them separately (sequential training in SL and concurrently in SFL) on a server and distributed BCI devices using their local brain data. As previously addressed in Subsection~\ref{sec:IEC}, training only a part of the neural network on the BCI device reduces its processing load and energy consumption, which is highly advantageous for resources-constrained devices. Besides, a BCI client has no access to the server-side model and vice-versa, providing better model privacy than FL~\cite{thapa2022splitfed}. Furthermore, the features of blockchain technology make it a robust solution for brain-related data protection. For example, the immutability of blockchain ensures that BCI data cannot be altered or deleted without proper authorization, maintaining the integrity of BCI records. As well, the distributed method of data storage adopted by the blockchain can effectively enhance the security of BCI data. In fact, brain signal samples will be stored across a decentralized network (multiple nodes) which make it more fault-tolerant than centralized approaches and more challenging for malicious nodes to compromise the entire data~\cite{khan2022blockchain}. 
However, despite the strong features offered by blockchain, its application in a BCI environment requires further investigation. For example, in literature there are numerous consensus algorithms with various performance and security properties, including Proof of Work (PoW), Proof of Stake (PoS), and Proof of Authority (PoA) to mention just a few~\cite{bamakan2020survey}. One consensus algorithm cannot satisfy the requirements of all BCI applications. Therefore, a technical comparison between the existing consensus algorithms within a BCI environment is needed. Additionally, it is important to note that the scalability issue of BCI blockchain systems becomes pronounced when dealing with large volumes of BCI data~\cite{sanka2021systematic}. From the above, we can say that (1) using standard security approaches alone is insufficient due to the unique properties of neural data and real-time control, (2) the selection of the right security method depends on the specific BCI use case, trade-offs, and the level of security and privacy required, (3) much further research efforts should be made to propose BCI-specific security solutions suitable for low-latency and privacy-sensitive applications.

\subsubsection{Realistic Scenario}

Consider a patient in a rehabilitation center operating a wheelchair or robotic limb through a BCI. The system continuously collects brain signals to interpret the user’s intentions. These neural signals are highly personal, reflecting goals, emotions, and cognitive states; therefore, protecting them is crucial. Transmitting such sensitive data over conventional networks poses risks of interception or tampering, potentially leading to privacy breaches, psychological profiling, or medical misuse.

A 6G-enabled BCI system integrated with edge computing can effectively mitigate these risks. A compact, high-performance edge server located within the rehabilitation facility processes the brain data locally and in real time, keeping it within secure institutional boundaries and minimizing exposure to cyber threats. Privacy-preserving measures such as hardware level encryption, local authentication, and biometric access control ensure that only authorized healthcare professionals can access processed outcomes. In this way, 6G and edge computing not only deliver ultra-fast and reliable neural processing but also uphold users’ mental privacy and ethical protection in BCI-based rehabilitation and assistive technologies.

\subsection{Zero Touch System Management}
\subsubsection{Introduction, Requirements and Potential Solutions}
Manual BCI systems management in 6G, including configuration, maintenance, troubleshooting, and software updates, is an inefficient, time-consuming, and error-prone. In~2017, the European Telecommunications Standards Institute (ETSI) developed the Zero Touch Network and Service Management (ZSM) framework to automate all processes and services, minimizing human intervention~\cite{etsi2019zero}. The framework utilizes emerging technologies such as AI, ML, DL, and Big Data to support network self-governing, including self-configuration, self-optimization, self-healing, and self-protection~\cite{benzaid2020ai}. The adoption of the ZSM concept in the context of 6G-enabled BCI can address many of these issues by automating various aspects of BCI applications and 6G networks. For example, ZSM can help automate the recognition process of 6G-connected BCI devices (types, specifications, and capabilities), using built-in algorithms and networking protocols. This feature helps the system to self-identify and self-configure different BCI devices without manual intervention. Machine and deep learning are among the key technologies used in ZSM frameworks to enable intelligent network management decisions~\cite{gallego2022machine}. ZSM, equipped with AI and machine learning algorithms, can automate the allocation of 6G network resources for BCI equipment. For example, if more BCI users and devices are connected, ZSM can recognize the need for additional 6G network resources and dynamically scale them (more bandwidth, computing resources, and routing capabilities) without manual configuration, ensuring a constant quality of service and positive user experience.
In the same way, BCI systems with ZSM features can self-analyze the BCI user's profiles such as age, gender, medical information, feedback reactions, usage goal, active hours, physical attributes (dominant hand, head size, and shape for EEG-based BCIs, etc.). This allows to automatically create a highly personalized BCI service that meets the BCI user's needs, comfort, and preferences without manual adjustments. In the same manner, the ZSM-driven 6G network can autonomously select optimal parameters for the success of a specific BCI operation or application, for example, high accuracy and precision in medical applications, while fast response time in a gaming context, ensuring efficient use of the network resource. As well, it can automate the creation of customized network slices for each specific BCI application~\cite{chergui2022toward}. Furthermore, AI-driven ZSM can assist in the ongoing surveillance of 6G-enabled BCI system for anomaly detection or network performance issues and implement self-diagnosis and self-corrective actions, reducing the need for human involvement. This process has significant value, especially in BCI-enabled healthcare settings where timely responses and uninterrupted service are critical. For example, consider an EEG-based BCI system within a 6G network that monitors patients' brain activity for early detection of neurological disorders, such as stroke or epilepsy. The incorporation of ZSM capabilities in the system permits a constant and remote monitoring of the EEG signals quality from patients. In this setting, ZSM employs highly trained deep learning algorithms to analyze EEG data in real time and discriminate between \textit{normal} and \textit{abnormal} brainwave patterns. If the system detects unusual EEG signals, it triggers an automatic self-generated response (e.g., bed alarm activation or even self-adjustment of treatment plans) to ensure the patient's safety. Similarly, in the event of network disruptions or failures, ZSM-enhanced 6G network can trigger self-healing mechanisms to restore connectivity and minimize downtime. ZSM can also automate the update process of BCI software. This ensures that BCI systems always run the latest and safest algorithms and features, improving their security and capabilities. Moreover, ZSM can contribute in enhancing the self-protection of sensitive brain-data by automating the detection and response to security threats and vulnerabilities. 

In light of all this, it is clear that ZSM plays an important role in improving the reliability, adaptability, scalability and security of 6G-enabled BCI systems. However, many limitations can impede the achievement of these objectives. Indeed, the high accuracy of ML/DL models in ZSM-enabled BCI in 6G networks is highly dependent on the availability of high-quality BCI-specific and 6G-specific datasets. Nevertheless, such datasets are scarce or even completely absent. Another issue is related to data volume, since the higher the volume of training data, the greater the potential for increased accuracy. Moreover, BCI models based on Supervised Learning (SL) or Semi-Supervised Learning (SSL) require annotated BCI data, which is a challenging task. In fact, annotating brain data necessitates the effort of an entire team of professionals with a deep understanding of brain structure, function, and neural pathways (neuroscientists, neurophysiologists, signal processing specialists, medical doctors, etc.). This can be difficult, time-consuming, and expensive, especially for large datasets. On top of that, most of ML/DL-enabled BCI models are considered as black-box models, meaning that they do not offer any explanations on how they reached their decisions. In this regard, the integration of Explainable Artificial Intelligence (XAI) approaches will enhance transparency, trust and effectiveness of intelligent BCI systems~\cite{guo2020explainable}. In addition, the use of BCI within a dynamic 6G network requires continuous adaptation of DL-based BCI models to the new data patterns. In this situation, MLOps practices~\cite{kreuzberger23} can streamline the retraining process, reduce the model adaptation time, and ensure that ZSM-based BCI systems perform optimally even in highly changing environments.

\subsubsection{Realistic Scenario} 

A patient with severe spinal cord injury uses a BCI-controlled wheelchair. The system includes a wearable unit that records brain signals and an application that translates them into control commands. When the device experiences a malfunction, the patient contacts a healthcare professional who accesses the BCI remotely through a Zero Touch System Management (ZTSM) platform. ZTSM employs machine learning to detect irregularities and provides real-time diagnostics, allowing the practitioner to identify and resolve the issue without physical access. The platform also ensures that the BCI receives regular updates and security patches, preventing future problems. Through ZTSM, device maintenance becomes faster, safer, and more efficient offering convenience to the patient and assurance to healthcare providers.

\section{6G-enabled BCI Use Cases} 
\label{sec:6G Use Case for BCI}
BCI technology is expected to enable novel human brain-connected systems interactions in everyday applications. Nevertheless, achieving this integration will necessitate certain network conditions. This section examines the constraints that previous mobile networks face in accommodating future BCI-enabled services and how the forthcoming 6G networks can address them.

\subsection{Digital Twin Brain}
\subsubsection{Motivation}

Many brain experiments cannot be performed directly on humans due to ethical, medical, and practical constraints. Real-time monitoring or stimulation of deep brain structures demands invasive techniques like craniotomy or depth electrode implantation~\cite{lozano2019deep, chen2024electrode}, which are limited to clinical cases and unjustifiable for healthy subjects. Likewise, procedures involving gene editing~\cite{ingusci2019gene}, brain region ablation~\cite{martinez2020randomized}, or artificial damage, though scientifically valuable, pose irreversible neurological and psychological risks, making them ethically unacceptable.

Consequently, researchers rely on animal models such as rodents, non-human primates, and zebrafish. For instance,~\cite{natoli2020does} used mice and primates to study COVID-19’s neurological effects, while~\cite{omar2023parkinson} induced Parkinson’s disease in zebrafish using MPTP injections. Similarly,~\cite{diehl2024bearded} explored Alzheimer’s-like pathology in aged capuchin monkeys. However, animal models cannot replicate uniquely human cognitive abilities, such as language comprehension, music perception, and theory of mind~\cite{kanwisher2025animal}. Therefore, to enable safe, precise, and ethical human brain experimentation, the development of human brain models becomes essential. In this context, Digital Twin (DT) technology, recognized as a core component of future 6G systems, can be integrated with BCIs, leading to the emergence of Brain Digital Twin (BDT) or Digital Twin Brain (DTB) applications.

\begin{figure}[!h]
\centering
  \includegraphics[width=\columnwidth, height=5.5cm]{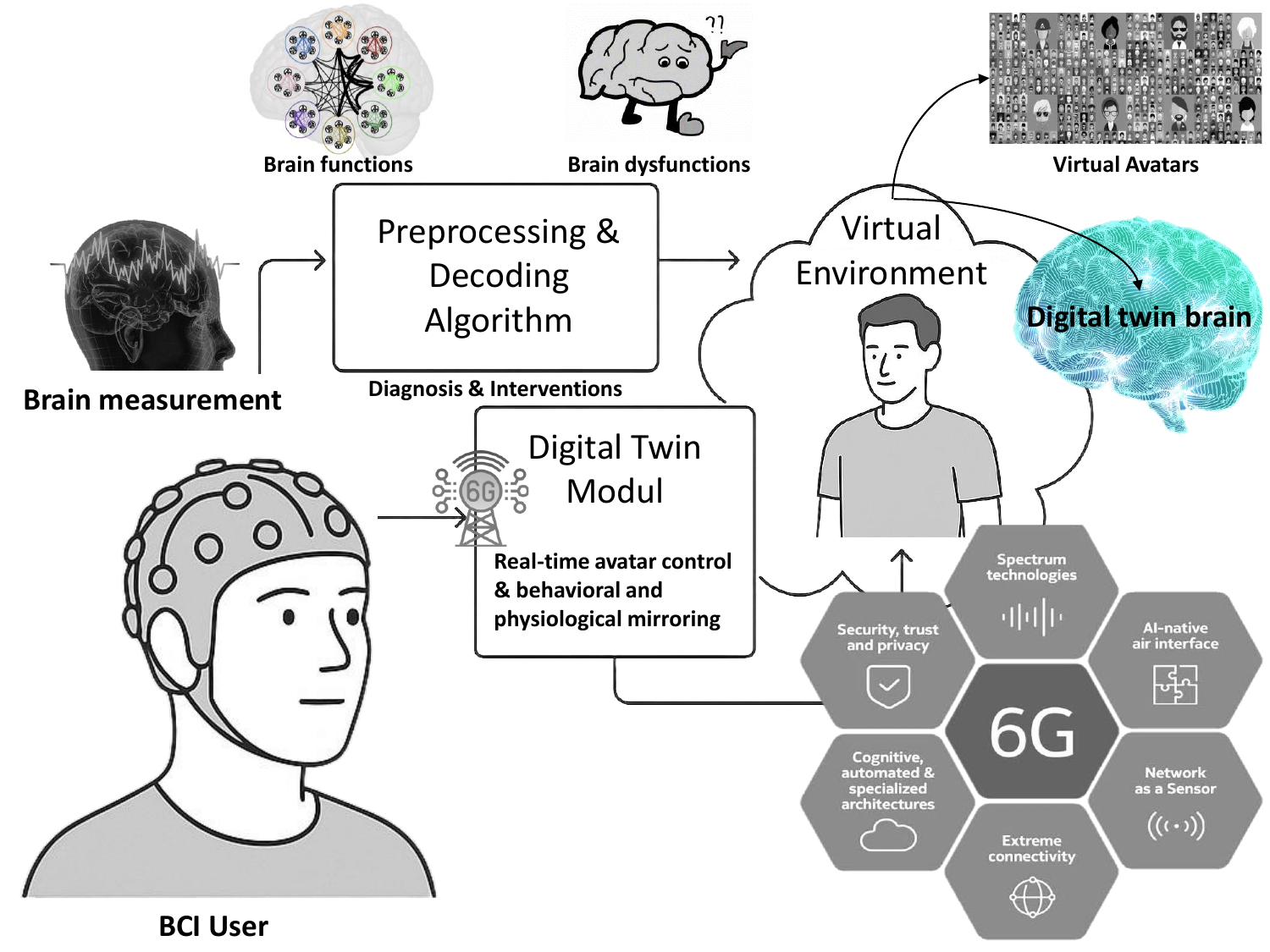}
  \caption{A 6G bridge between biological and artificial intelligence to create a digital twin brain.}
   \label{Scenario2}
\end{figure}

BDT or DTB can bridge the gap between biological and artificial intelligence in the form of brain networks. The concept of a digital twin was initially introduced by NASA during the Apollo mission in the United States ~\cite{xiong2023digital}. It involves modeling virtual replicas of the real brain that would, at all times, mirror the most recent features of the physical brain, allowing continuous remote monitoring throughout its entire lifetime ~\cite{fekonja2024digital}. It can be classified into three categories, namely, brain digital monitoring twin, brain digital simulation twin, and brain operational twin~\cite{khan2022digital}. A monitoring brain twin enables the monitoring the status of the real brain, whereas a simulation brain twin uses various simulation tools and machine learning schemes to provide insights about future brain states. Meanwhile, an operational brain twin enables neurologists and neurosurgeons to interface with the physical brain and perform different neurological interventions in addition to analysis and system modeling. 

The literature is not yet rich in this area, and even the few studies that have explored the topic fall short of achieving a complete, dynamic, and real-time brain digital twin system. For example, in~\cite{wan2021semi}, an improved AlexNet model and semi-supervised support vector machines (S3VMs) are used within a digital twin environment to develop a diagnostic and predictive framework for brain image processing. The framework maps real brain images into a virtual environment using MRI data, enabling accurate brain image feature recognition and digital diagnosis. More recently, to support early diagnosis and treatment in brain tumor management, the authors in~\cite{sarris2023towards} attempted to develop a personalized virtual replica of a patient's brain by integrating 3D MRI images, EEG recordings, and real-time IoT sensor data. The framework combines a K-means-based brain tumor detection algorithm with a 3D brain model. Both studies remain static and offline systems. They lack the dynamic, real-time data integration, continuous monitoring, and bi-directional interaction required for a fully functional brain digital twin.

\subsubsection{How 6G can help}
Three core components constitute the digital brain twin: the physical brain, the virtual brain, and the AI-processed, command, and control from the virtual brain to the real brain. The communication part plays an important role in the successful synergy between the brain twin model and its corresponding. In such a scenario, network reliability is extremely important, and a fully wired connection is not a viable solution due to its complexity of installation and high cost. Compared to 5G, where approximately eight and a half hours of annual downtime and a reliability of 99.9\%, 6G promises a realiability of 99.999\% and just 5 minutes of downtime per year ~\cite{ahmadi2022networked}. Moreover, to reach the full potential of brain digital twin, an ultra-low-latency communication between the physical brain and the digital counterpart is non-negotiable. For example, in a brain digital twin predictive model, a high latency can result in a desynchronization loop between the physical brain and its digital replica. In such a case, the brain digital twin processes outdated neural signals that no longer represent the current state of the biological brain, leading to wrong decisions (e.g., trigger false alarms or miss real danger). In addition, a user experiencing a consistent delay between intent and response will quickly lose confidence in the twin system and possibly reject the use of the interface in the future. In response, 6G reduces latency by 10x to 100x compared to 5G, making it critical for real-time brain-digital twin interactions, where sub-millisecond delays are essential for accurate neural feedback.  

The transition to 6G is not merely an incremental progression but a fundamental paradigm shift in mobile communication technology. Unlike the previous generations, which were primarily about higher data rates and enhanced connectivity, 6G is being envisioned as an intelligent, immersive, and adaptive network fabric that synergistically converges the physical, digital, and biological worlds. This infrastructure of the future will enable next-generation BCI applications such as neural dust—ultra-miniaturized, implantable wireless sensors that use ultrasonic power and communication for biological tissue interfaces ~\cite{warneke2001smart}. Developed at the University of California, Berkeley, these micrometer-scale devices enable real-time recording of brain signals for applications such as prosthetic control, neuromodulation, and organ diagnostics—all demanding gigabit-per-second data rates, sub-millisecond latency, and high reliability. In addition, one of the 5G bottlenecks in maximizing the potential of brain digital twin is its network speed. For example, in a brain-digital twin system that streams high-resolution neural activity from a subset of the visual cortex (about 1 billion neurons) to an edge server in real time, if we assume an average spike rate of 10–100 Hz and around 100 bits of data per spike, the system would generate a data rate of approximately 100 Gbps. While 5G can only offer around 1 to 3 Gbps per user with higher latency, 6G targets 100 to 200 Gbps, and sub-millisecond latency makes it the viable option for this level of neural data streaming. Moreover, with network slicing, brain-twin traffic can receive dedicated resources, facilitating the transmitting of high-resolution brain signals (EEG, fMRI, etc.).

The human brain is highly complex, characterized by non-linear, dynamic, and spatio-temporal interactions with multimodal, noisy, and high-dimensional data. Modeling such systems in real time requires effective machine and deep learning approaches. However, centralized training of brain digital twin models is often impractical due to latency, bandwidth limitations, and security or privacy risks. To mitigate these issues, 6G will integrate distributed intelligence at the network edge, allowing multiple brain digital twin models to be trained locally and then aggregated centrally through iterative updates until convergence. Although distributing computation reduces training time, it increases communication overhead, reaching a saturation point beyond which additional machines do not improve overall efficiency~\cite{wen2017terngrad}. Sparse federated learning can address this by transmitting only significant gradient updates, reducing communication costs and improving scalability~\cite{kim2024spafl}.

Despite its benefits, conventional federated learning remains vulnerable to failures or attacks on the central aggregator. Dispersed federated learning with distributed aggregation offers greater robustness by removing single-point dependencies~\cite{khan2021dispersed}. Moreover, brain activity is non-stationary and personalized, rendering static offline learning ineffective. Thus, online learning approaches are essential for maintaining adaptive, real-time brain digital twins capable of tracking continuous brain dynamics—critical in health and assistive contexts~\cite{hoi2021online}. Beyond computation, 6G will support holographic and extended reality (XR) communication, enabling immersive visualization of brain states. For instance, a neurosurgeon could employ augmented reality goggles to visualize a patient’s digital twin in 3D during surgery, guided by real-time predictive brain models transmitted over 6G.

Although digital twin technology already exists in some industrial applications supported by 5G or even 4G~\cite{kritzinger2018digital}, it has not reached its full potential in the BCI. The integration of 6G and brain digital twin technology is poised to revolutionize healthcare, neuroscience, and neuro-AI integration by enabling real-time, ultra-reliable, and intelligent communication between digital and biological brain systems. 

\subsection{Brain Immersion Systems}
\subsubsection{Motivation}
Immersive technologies, including VR, AR, and MR, introduced groundbreaking solutions in various sectors such as education, gaming, medicine, marketing, and robotics~\cite{suh2018state}. For example, in education, virtual reality allows learners to interact with complex concepts in 3D and explore inaccessible environments (e.g. inside a volcano, ancient cities, solar system)~\cite{rojas2023systematic}. In robotics, mixed reality enables users to visualize and interact with a digital twin of a robotic arm superimposed on the real world. This immersive interface allows for precise and efficient robot control and task programming in robotic-assisted surgery or industrial automation~\cite{yu2022mr}. 

However, despite their popularity and technical relevance, most of these technologies, particularly VR and MR, are often based on handheld controllers that can challenge new or inexperienced users and reduce immersion. The level of immersion in virtual environments has a great impact on the user's experience. A highly immersive environment increases the sense of realism and presence, while a less immersive space leads to distraction and lowers user satisfaction.  In addition, these devices require a certain level of manual adroitness and hand-eye coordination, which can pose effective use challenging or even impossible for people with physical disabilities. According to the World Health Organization, approximately 1.3 billion people (about 16\% of the global population) experience significant disabilities~\cite{WHO}. Nevertheless, immersive systems research and development rarely consider these individuals, leading to exclusive and inaccessible experiences. For example, players with lower limb MI disabilities may find it very difficult and unsafe to perform various locomotor movements in a virtual museum tour game without fear of falling or injuring themselves due to balance problems, lack of wheelchair support, and risk of motion sickness. Likewise, students with vision loss cannot benefit from visual-heavy environments, while those with auditory disabilities may miss spoken instructions or ambient cues. In addition, learners with motor impairments can struggle with standard VR controllers, which often require motor skills and fast reflexes. This can result in emotional and psychological disturbance. 

The integration of BCI would ensure equal opportunities for all users, allowing not only able-bodied individuals but also those with physical disabilities to control their favorite VR-based video games with only their minds, reducing the reliance on physical devices. First, EEG headsets detect brain activity related to attention, motor imagery, or specific cognitive states. These signals are then processed and translated into in-game commands (e.g., move forward, select an object, fire a weapon). Finally, the BCI system is integrated with VR platforms through software APIs, allowing more natural thought-based interaction. This can significantly improve immersion level by bridging the gap between the player's intentions and in-game actions~\cite{wozniak2021enhancing}. A further example can be seen in a BCI-enabled VR horror game, where the level of terror, anxiety, and intensity can be dynamically adjusted in real-time based on the user's dread level. 

In virtual classrooms, the incorporation of BCI helps students with MI or communication difficulties engage in discussions, respond to questions, and complete verbal or written activities, translating their neurological signals into text or speech~\cite{soman2015using}. This would empower independence and equality in educational access, inspire disabled students to express their ideas, and participate in immersive classrooms.  Moreover, BCI can significantly enhance learning immersion by monitoring students' focus and attention through real-time detection of exhaustion or distraction. Based on the cognitive state of the student, the system can automatically adjust the difficulty level and pace of the lesson, suggest breaks, or change to lighter tasks, helping students stay mentally present, which is a central feature of immersive experiences~\cite{peng2021fatigue}. 

Some studies in the literature have already explored the incorporation of BCI, either independently or alongside existing immersive technologies, to enhance the user's immersive experience. For example, in~\cite{li2021mindgomoku}, the authors developed the MindGomoku game (BCI version of the Gomoku game), providing a practical and accurate approach to controlling the game using EEG signals, through a P300-based BCI system, making the experience accessible, particularly for disabled users. The authors in~\cite{wang2023portable} developed and evaluated an augmented reality-based BCI system that allows stroke patients to control a rehabilitation exoskeleton using only their brain signals. In~\cite{cheng2023haptic}, the authors proposed a method to improve the quality of MI EEG signals for virtual reality BCI applications by using haptic stimulation training to enhance signal classification and control accuracy. The study conducted in~\cite{DENG2023104944}, developed a VR-based BCI system for intelligent control of UAV swarms, enabling users to control multiple drones using brain signals through an immersive and flexible interface. Recent research in~\cite{PanConf} developed a personalized VR-based Mindfulness-Based Cognitive Therapy (MBCT) system that uses real-time EEG data via BCI to recognize and reduce exam anxiety in students through emotion-adaptive interventions.

\subsubsection{How 6G can help} 
Brain immersion systems require different performance metrics compared to what 5G delivers. A concrete example can be found in~\cite{gehrke2025neuroadaptive}, where the authors investigated a neuroadaptive haptic system that integrated EEG-based BCI with visual VR and haptic feedback, using reinforcement learning to adjust haptic intensity based on real-time EEG signals. This setup requires the simultaneous transmission of high-bandwidth visual data, ultra-low-latency haptic commands, and continuous neural feedback, a combination that exceeds the capabilities of current 5G networks. 5G lacks the very low latency and bandwidth coordination to handle all these multimodal inputs in real time~\cite{saad2019vision}, highlighting the need for more advanced connectivity solutions.

6G communication guarantees ultra-low-latency (sub-ms), essential for real-time haptic feedback and neural response adaptation. In addition, leveraging higher spectrum technologies such as Terahertz (THz) and Non-RF (laser or visible light communication), 6G will provide an additional 1000x increase in data rates (Tbps-level speeds)~\cite{de2021survey}. This will support simultaneous high-resolution visual streams, multi-channel EEG data, and haptic signals.  Moreover, unlike 4G systems, which lacked AI involvement, and 5G communication systems, which are expected to offer only partial or limited AI support, 6G networks are expected to be fully driven by AI for comprehensive automation. 6G will integrate Multi-Access Edge Computing (MEC) with advanced AI algorithms, enabling intelligent decoding of EEG signals at the network edge to personalize and adapt the XR experience in real time. 

Furthermore, future 6G networks are expected to shift network design from the traditional Shannon framework to goal-oriented semantic communication~\cite{strinati20216g}. Instead of forwarding the entire raw data to a server or cloud for processing, which requires massive amount of bandwidth and introduces high transmission time, the emerging philosophy of next-generation networks is the implementation of distributed computing mechanisms capable of learning and extracting meaning from data, exploiting proper knowledge representation systems, and identifying strictly relevant information in goal-oriented communications. For example, in~\cite{zheng2024eideticomcrossmodalbraincomputersemantic}, the authors presented EidetiCom, a BCI framework that considers the brain as a semantic source. The idea consists in compressing EEG to capture meaningful information, such as visual categories or captions, and transmitting only the decoded semantic intent, rather than the full neural data. The research illustrates how future brain systems can use a semantic-level communication layer, reducing bandwidth usage, and focusing on user intent.

Importantly, to fully measure or improve the immersion level of the user, traditional network metrics such as QoS (e.g., latency, data rate, and packet loss) and QoE (e.g., mean opinion score) are not enough. The study carried out in~\cite{kasgari2019human}, demonstrated that the human brain may not be able to detect differences in latency measures, within the URLLC range. Meanwhile, the same authors revealed in~\cite{park2019wireless}, that visual and tactile inputs play an important role in optimizing resource efficiency. Therefore, in addition to the classical indicators, 6G will incorporate a new user-centric performance metric, aka Quality of Physical Experience (QoPE). This metric merges the perceptual, cognitive, and physiological factors of humans in the performance evaluation of the brain-driven immersive system~\cite{saad2019vision}. ~\tablename~\ref{tab: brain immersion systems} presents an in-depth review of recent works tha combine immersive technologies with BCI and illustrates the role of 6G in their advancements.

\begin{table*}[]
\centering
\caption{Analysis of 6G contributions in brain immersion systems studies.}
\label{tab: brain immersion systems}
\resizebox{\textwidth}{!}{%
\begin{tabular}{@{}llllll@{}}
\toprule
\textbf{Work, Year, Ref} &
  \textbf{\begin{tabular}[c]{@{}l@{}}BCI \\ Paradigm\end{tabular}} &
  \textbf{Target Application} &
  \textbf{Key Findings} &
  \textbf{Limitations} &
  \textbf{6G Enhancements} \\ \midrule
\begin{tabular}[c]{@{}l@{}}Du et al.,\\ 2022,~\cite{du2022visual}\end{tabular} &
  SSVEP &
  \begin{tabular}[c]{@{}l@{}}AR-based BCI using four colors\\ (white, red, green, blue) displayed\\ on a computer and HoloLens to test\\ color effects on BCI performance.\end{tabular} &
  \begin{tabular}[c]{@{}l@{}}- Color affects AR-SSVEP and PC-SSVEP \\ differently.\\ - Green are best for shorter durations ($<$ 1.5) \\ in AR-SSVEP\\ - Red \& white are best for longer durations \\ ($>$1.5 s)  in AR-SSVEP\\ - Red yield the highest ITR in PC-SSVEP\\ - Blue is the least effective in both systems\end{tabular} &
  \begin{tabular}[c]{@{}l@{}}- Limited to 2D despite 3D space, reducing \\ user engagement.\\ - The study focuses on color and duration,\\ excluding other factors.\end{tabular} &
  \begin{tabular}[c]{@{}l@{}}- Transmission of high-quality 3D visuals, \\ for immersive and more complex \\ AR-BCI experiences.\\ -Real-time edge processing of brain data,\\ reducing the computational load on HoloLens devices\\ - Integrate multi-modal data (such as tactile and \\ auditory inputs)\end{tabular} \\ \midrule
\begin{tabular}[c]{@{}l@{}}Wang et al.,\\ 2023,~\cite{wang2023portable}\end{tabular} &
  SSVEP &
  \begin{tabular}[c]{@{}l@{}}AR-BCI system for rehabilitation \\ exoskeletons\end{tabular} &
  \begin{tabular}[c]{@{}l@{}}- Offline recognition accuracy: 90.2\% \\ (2.5 s, $>$90\% with Oz and O2)\\ - Online classification accuracy: 88.9\%\\ ITR: 30.01 bit/m\\ - Exoskeleton movement accuracy: 91.12\%\\ ITR: 31.63 bits/m\end{tabular} &
  \begin{tabular}[c]{@{}l@{}}- AR-BCI performance is still limited \\ compared to computer screen BCI.\\ - Focused on electrode setup, neglecting\\ user variability and environement factors\end{tabular} &
  \begin{tabular}[c]{@{}l@{}}- Integration of more wearable devices \\ and sensors for better patient monitoring.\\ - Transmission of a large amount of brain data \\ from multiple electrodes\\ - Real-time feedback enhances stroke rehab \\ responsiveness\end{tabular} \\ \midrule
\begin{tabular}[c]{@{}l@{}}Deng et al.,\\ 2023,~\cite{DENG2023104944}\end{tabular} &
  SSVEP &
  \begin{tabular}[c]{@{}l@{}}UAV swarm control via BCI \\ based on VR.\end{tabular} &
  \begin{tabular}[c]{@{}l@{}}- Control accuracy of quadcopter clusters: 90\%\\ - Average ITR: 3.62 bit/min (online)\end{tabular} &
  \begin{tabular}[c]{@{}l@{}}- Wireless network conditions can bottleneck \\ large UAV swarms.\end{tabular} &
  \begin{tabular}[c]{@{}l@{}}- 6G can handle large-scale UAV swarms with \\ high-speed data transmission.\\ - 6G can reduce the communication delays.\\ - Faster data rates for VR and EEG transmission\end{tabular} \\ \midrule
\begin{tabular}[c]{@{}l@{}}Cheng et al.,\\ 2023,~\cite{cheng2023haptic}\end{tabular} &
  MI &
  \begin{tabular}[c]{@{}l@{}}BCI in VR for controlling ball \\ movement with EEG signals.\end{tabular} &
  \begin{tabular}[c]{@{}l@{}}- Improved MI EEG quality with haptic training: \\ Left-hand differentiation +21.8\%,\\ Right-hand +15.7\%; \\ Classification accuracy at 93.5\%.\end{tabular} &
  \begin{tabular}[c]{@{}l@{}}- Discomfort reported by participants wearing\\  both VR headset and EEG cap\end{tabular} &
  \begin{tabular}[c]{@{}l@{}}- Reduces lag, enhances immersion\\ - Improves response time for haptic feedback\\ - Integrates lightweight EEG and VR devices\end{tabular} \\ \midrule
\begin{tabular}[c]{@{}l@{}}Tezza et al.,\\ 2020,~\cite{tezza2020brain}\end{tabular} &
  MI &
  \begin{tabular}[c]{@{}l@{}}Brain-controlled drone racing \\ video game\end{tabular} &
  \begin{tabular}[c]{@{}l@{}}- 50 of 54 participants expressed willingness \\ to buy a BCI game for $200 to $500.\\ - High excitement among participants\\ - Preferred game genres: FPS, racing,\\  MMORPG\end{tabular} &
  \begin{tabular}[c]{@{}l@{}}- Brain signals can be noisy and affected \\ by external factors, causing inaccuracies \\ in drone control.\\ - Connectivity issues\end{tabular} &
  \begin{tabular}[c]{@{}l@{}}- Ultra-reliable low-latency communication, \\ for real-time control in drone racing games.\\ - Greater data transfer rate\\ - Complex BCI data processing\\ - high Quality VR integration\end{tabular} \\ \midrule
\begin{tabular}[c]{@{}l@{}}Al et al.,\\ 2022,~\cite{al2022predict}\end{tabular} &
  \begin{tabular}[c]{@{}l@{}}Passive \\ monitoring\\ of EEG\end{tabular} &
  \begin{tabular}[c]{@{}l@{}}BCI system for monitoring student \\ attention in online classes.\end{tabular} &
  \begin{tabular}[c]{@{}l@{}}- 96\% accuracy with random forest (RF).\\ - RF outperformed KNN and SVM\end{tabular} &
  - Focused on three classifiers only. &
  \begin{tabular}[c]{@{}l@{}}- 6G's low latency enables real-time\\ feedback to both students and instructors\\ - High-speed EEG transfer from multiple students\\ - Comprehensive attention analysis across a larger\\ group of learners\\ - Combine EEG with other biometric signals\\ due to the seamless connectivity of multiple sensors\\ and IoT devices\end{tabular} \\ \midrule
\begin{tabular}[c]{@{}l@{}}Armani et al.,\\ 2023,~\cite{armani2023maths}\end{tabular} &
  \begin{tabular}[c]{@{}l@{}}Passive \\ monitoring\\ of EEG\end{tabular} &
  \begin{tabular}[c]{@{}l@{}}EEG-based adaptive BCI \\ for math learning: investigate \\ cognitive states in math anxiety\end{tabular} &
  \begin{tabular}[c]{@{}l@{}}- Math-anxious individuals solve problems faster.\\ - Performance remains stable without \\ overload/distraction\\ - No click difference between Math-anxious \\ and non-anxious participants\end{tabular} &
  \begin{tabular}[c]{@{}l@{}}- Small sample size (10 participants)\\ - Impact of distractions and memory overload\\ not fully explored\end{tabular} &
  \begin{tabular}[c]{@{}l@{}}- Integration of AR/VR into BCI, \\ creating immersive learning \\ that engage students and reduce Math Anxiety.\\ - 6G's high connectiviy enables remote participation,\\ enhancing diveristy and sample size\end{tabular} \\ \bottomrule
\end{tabular}%
}
\end{table*}

\subsection{Brain-Artificial Intelligence Interface for Assistive Healthcare}
\subsubsection{Motivation}

The risk of developing brain and spinal cord disorders, nerve damage, and limb loss increases significantly, particularly among the elderly~\cite{belkacem2020brain}. Such conditions negatively affect patients’ quality of life. For instance, spinal cord injury (SCI) leads to reduced mobility and autonomy, limiting basic daily activities such as walking or standing~\cite{bryce2021spinal}. Likewise, Parkinson’s disease (PD) causes progressive motor decline due to dopaminergic neuron degeneration, resulting in tremor, rigidity, bradykinesia, and speech difficulties~\cite{coleman2020rethinking}. Collectively, these symptoms affect physical, psychological, and social well-being, often leading to frustration, depression, and loss of independence.

Various Assistive Technologies (ATs) have been developed to support individuals with disabilities, including prosthetics, wheelchairs, screen readers, cochlear implants, and speech-generating devices. However, many people with severe motor impairments still cannot fully benefit from these solutions. For example, users with limited arm or hand mobility may struggle to control traditional wheelchairs, while conventional prosthetics rely on residual muscle movements that are tiring and imprecise. Moreover, most prosthetics lack sensory feedback, preventing users from perceiving touch, temperature, or pressure, and making interactions less natural~\cite{jabban2022sensory}. These limitations highlight the need for intelligent, adaptive systems that can address the diverse requirements of disabled individuals.

Integrating BCI technology into assistive systems offers a promising path forward. BCIs can enable users to control wheelchairs or robotic limbs solely through neural activity~\cite{naser2023towards}, significantly reducing physical effort. They can also extend to smart home environments, allowing users to turn on devices, open doors, or adjust temperature through thought commands, creating a seamless and muscle-free interaction between the brain and external systems.

Depending on the BCI technique, different implementations have been demonstrated. For example,~\cite{cortez2020smart} proposed a P300-based BCI for post-stroke patients to control home appliances using Multilayer Perceptron (MLP) and Support Vector Machine (SVM) classifiers. In~\cite{saboor2017ssvep}, a hands-free system controlled light switches and appliances via SSVEP stimuli on smart glasses. Similarly,~\cite{chai2020hybrid} introduced a hybrid SSVEP–EMG BCI for paralyzed users. EEG-based BCIs have also enabled control of prosthetic fingers~\cite{gannouni2020eeg}, phantom limb pain relief~\cite{yanagisawa2019using}, and prosthetic leg movement~\cite{constantine2021bci}. These systems enhance users’ independence and quality of life (QoL), allowing more natural, real-time interaction with their environment. Moreover, sensory-enabled BCI prosthetics can restore tactile feedback for improved perception~\cite{fleury2020survey}.

A new generation of Brain–Artificial Intelligence Interfaces (BAIs) extends traditional BCIs by integrating artificial intelligence to replace parts of the neurocognitive process. BAIs enable users to express high-level intentions, which are translated into actions by an AI agent, thereby benefiting individuals with cognitive impairments often excluded from standard BCI applications. For example,~\cite{meunier2024conversational} introduced a Conversational BAI driven by EEG that simulates phone conversations, enabling meaningful, fluent communication without language generation. This marks a major advance in speech neuroprosthetics, demonstrating the feasibility of non-invasive systems for natural interaction in real-world scenarios.

\subsubsection{How 6G can help}
Although 5G technology supports certain healthcare use-cases~\cite{ahad20195g}, there are still specific applications where their requirements exceed the capabilities of 5G networks. For example, real-time BCI surgery, such as Deep Brain Stimulation (DBS) surgery for Parkinson's disease~\cite{brown2016controlling} where the BCI system should accurately record, process, and analyze brain activity patterns, enabling physicians to deliver targeted stimulation to the appropriate brain regions at the appropriate time (in conditions like epilepsy) and with high precision. In such cases, any errors or delays, even of microseconds, can significantly impact patient safety and treatment success. Factors such as network congestion, signal interference, and variation in signal strength can affect the latency and reliability of BCI data transmission. BCI-enabled monitoring systems represent another scenario that requires dedicated and highly reliable communication channels to ensure continuous data acquisition and real-time analysis, particularly in environments with high levels of electromagnetic interference or network congestion. 

The use of Tera-Hz frequencies in 6G networks enables ultrafast transmission of large brain signals for timely diagnoses. Physicians can continuously monitor accurate EEG signals and deploy a healthcare solution remotely. High-speed connectivity also ensures minimal latency and reliable transmission, essential for remote monitoring and telemedicine applications.  This can be especially advantageous for patients with limited mobility or in remote areas, as it reduces hospital visits. For example, during pandemic outbreaks such as COVID-19, a mental 6G-enabled health application with BCI integration can provide real-time monitoring of patients' neurological signals to assess their mental and emotional states. By analyzing brain activity patterns associated with stress, anxiety, or other mental health conditions, the system can offer personalized interventions and support patients remotely, minimizing the risk of disease transmission.

Holographic communication and virtual reality are also among the 6G-driven innovations that will contribute to the advancement of BCI-enabled intelligent healthcare systems. As addressed earlier, both technologies will need to transfer much larger amounts of data, at much higher speeds for optimal performance. 5G cannot meet the complex and increasing needs of future healthcare applications. The next generation of communication technology, 6G, is expected to deliver such a high level of QoS. 

6G can improve the capabilities of BCI systems by making communication between the user's brain signals and the devices they wish to control both faster and more reliable. With the high-speed and low-latency connectivity offered by 6G networks, communication between the brain of a patient and a prosthetic device or exoskeleton can be carried out with greater precision, thus improving the overall performance of the prosthetic device~\cite{saad2019vision}. A person with paralysis can manage a smart home system through BCI (turn on lights, regulate the thermostat, and even unlock doors) in real-time, ensuring instant feedback and control.

BCI applications demand highly specialized resources, including memory, processing, and network resources, which can be challenging to manage. The synergy between artificial intelligence, edge computing, and 6G technology will bring several benefits to deal with these issues. Cloud capabilities will be closer to the point of care, enabling local storage, processing, and analysis of massive BCI-generated data. This facilitates faster insights into brain activity patterns, allowing for more accurate diagnoses and personalized treatment plans~\cite{nayak20216g, mucchi20206g}. Edge computing can also minimize the amount of data that must be transferred over the network, which is particularly advantageous for distant or mobile BCI applications. Meanwhile, cloud computing can supply the high-performance computer resources required to process the vast data generated by BCI applications. 
Furthermore, 6G networks can facilitate the application of machine learning methods to enhance prosthesis control. Machine learning may be used to examine data from prosthesis sensors and user data to enhance control algorithms and increase movement precision. ~\tablename~\ref{tab: assistive healthcare} summarizes key studies on brain-driven assistive healthcare and illustrates how 6G can support their evolution.

\begin{table*}[!h]
\centering
\caption{Analysis of 6G contributions in brain-AI interface for assistive healthcare studies.}
\label{tab: assistive healthcare}
\resizebox{\textwidth}{!}{%
\begin{tabular}{@{}llllll@{}}
\toprule
\textbf{Work, Year, Ref} &
  \textbf{\begin{tabular}[c]{@{}l@{}}BCI \\ Paradigm\end{tabular}} &
  \textbf{Target Application} &
  \textbf{Key Findings} &
  \textbf{Limitations} &
  \textbf{6G Enhancements} \\ \midrule
\begin{tabular}[c]{@{}l@{}}Bairagi et al.,\\ 2018, ~\cite{bairagi2018eeg}\end{tabular} &
  EEG &
  Early diagnosis of Alzheimer’s disease &
  \begin{tabular}[c]{@{}l@{}}- Diagnostic accuracy: 94\%\\ (combination of spectral and wavelet features)\end{tabular} &
  - Reliance on the quality of EEG data &
  \begin{tabular}[c]{@{}l@{}}- Advanced signal processing for\\  better feature extraction\\ - Facilitate remote monitoring and diagnosis\\ through high-speed connectivity\end{tabular} \\ \midrule
\begin{tabular}[c]{@{}l@{}}Chai et al.,\\ 2020, ~\cite{chai2020hybrid}\end{tabular} &
  \begin{tabular}[c]{@{}l@{}}SSVEP \\ + \\ EMG\end{tabular} &
  \begin{tabular}[c]{@{}l@{}}Smart home control system\\ for individuals with paralysis\end{tabular} &
  \begin{tabular}[c]{@{}l@{}}- Target selection acc:\\   - 97.5\%  (healthy), 83.6\% (paralysis)\\ - Confirmation acc:\\   - 97.6\% (healthy),  96.9\% (paralysis)\\ - Maximized acc: 100\% (SSVEP + EMG)\\ - Information Transmission Rate: 45bit/m\end{tabular} &
  \begin{tabular}[c]{@{}l@{}}- Participant diversity: only 5 healthy and \\ 5 individuals with paralysis\\ - Authors use only a single-channel \\  EMG\end{tabular} &
  \begin{tabular}[c]{@{}l@{}}- High data rates for transmission of multiple\\ EEG and EMG signals.\\ - Reliable connections among the different \\ smart home devices.\\ - 6G supports massive connectivity, allowing\\ seameless control of multiple smart home devices\\ simultaneouly.\end{tabular} \\ \midrule
\begin{tabular}[c]{@{}l@{}}Sebastian et al.,\\ 2020, ~\cite{sebastian2020brain}\end{tabular} &
  MI &
  \begin{tabular}[c]{@{}l@{}}Enhancing stroke rehabilitation\\ using BCI, VR and FES\end{tabular} &
  - MI accuracy $>$ 80\% &
  \begin{tabular}[c]{@{}l@{}}- Low processing speed and signal \\ reliability, especially in noisy environments\\ - Latency in VR and FES feedback\end{tabular} &
  \begin{tabular}[c]{@{}l@{}}- Scalable and real-time remote monitoring\\ for stroke patients\\ - Improve the VR and FES systems' responsiveness\end{tabular} \\ \midrule
\begin{tabular}[c]{@{}l@{}}Cruz et al.,\\ 2021, ~\cite{cruz2021self}\end{tabular} &
  P300 &
  \begin{tabular}[c]{@{}l@{}}Brain-controlled wheelchairs \\ (BCWs)\end{tabular} &
  \begin{tabular}[c]{@{}l@{}}- Driving accuracy:\\    - Able-bodied participants: 95.8\%\\    - Physically disabled participants 93.7\%\\    - $>$ 99\% after corrections from\\       collaborative control\end{tabular} &
  \begin{tabular}[c]{@{}l@{}}- Limited sample size (13 in total)\\ - Limited scenario variability (office)\\ such as crowded spaces and outdoor navigation.\end{tabular} &
  \begin{tabular}[c]{@{}l@{}}- Advanced sensor integration (e.g, HD cameras\\ and muliple IoT devices) in BCWs\\ - Scale the BCWs to operate in multi-user and complex\\ scenarios without losing performance (network slicing)\\ - Make BCWs more responsive\end{tabular} \\ \midrule
\begin{tabular}[c]{@{}l@{}}Dillen et al.,\\ 2022, ~\cite{dillen2022data}\end{tabular} &
  MI &
  Control lower limb prosthetics &
  \begin{tabular}[c]{@{}l@{}}- Decoding accuracy: 84\% for \\ both groups (limb amputation and \\ able-bodied)\end{tabular} &
  \begin{tabular}[c]{@{}l@{}}- EEG data input included 2-second \\ post-movement delay\\ - Lack of real-time interaction\\ - The study focuses only on single-joint\\ movements\end{tabular} &
  \begin{tabular}[c]{@{}l@{}}- 6G-enabled edge EEG processing reduces latency \\ and improves prosthetic responsiveness\\ - Advanced ML/DL methods for improved\\ neural modeling in prosthetics\\ - 6G's high bandwidth enables to process large and\\ complex data, decoding multiple movements\end{tabular} \\ \bottomrule
\end{tabular}%
}
\end{table*}

\subsection{Brain-Net and Internet of Minds} 
\subsubsection{Motivation}
Researchers have been exploring ways to connect multiple brains to create a Brain Network (BrainNet) system. This system is direct brain-to-brain interface (B2BI) for collaborative problem-solving (see~\figurename~\ref{Scenariobrainnet}). It was demonstrated using non-invasive B2BI which combines noninvasive brain measurement (e.g., EEG) to record brain signals and brain stimulation technique (e.g., Transcranial magnetic stimulation, TMS) to noninvasively transmit brain commands or information. In study \cite{jiang2019brainnet}, three participants—two Senders and one Receiver—collaborate in a Tetris-like game. The Senders’ EEG signals are analyzed in real-time to determine their decisions about rotating a block, which are then transmitted to the Receiver’s brain via TMS. The Receiver, who cannot see the screen, integrates this input and makes a decision using an EEG interface. A second round allows for feedback and potential corrections. Additionally, experiments showed that the receiver learns to trust the more reliable sender, similar to social networks. This research paves the way for future communication networks for cooperative B2BI tasks.

\begin{figure}[!h]
\centering
  \includegraphics[width=1\linewidth, height=5.5cm]{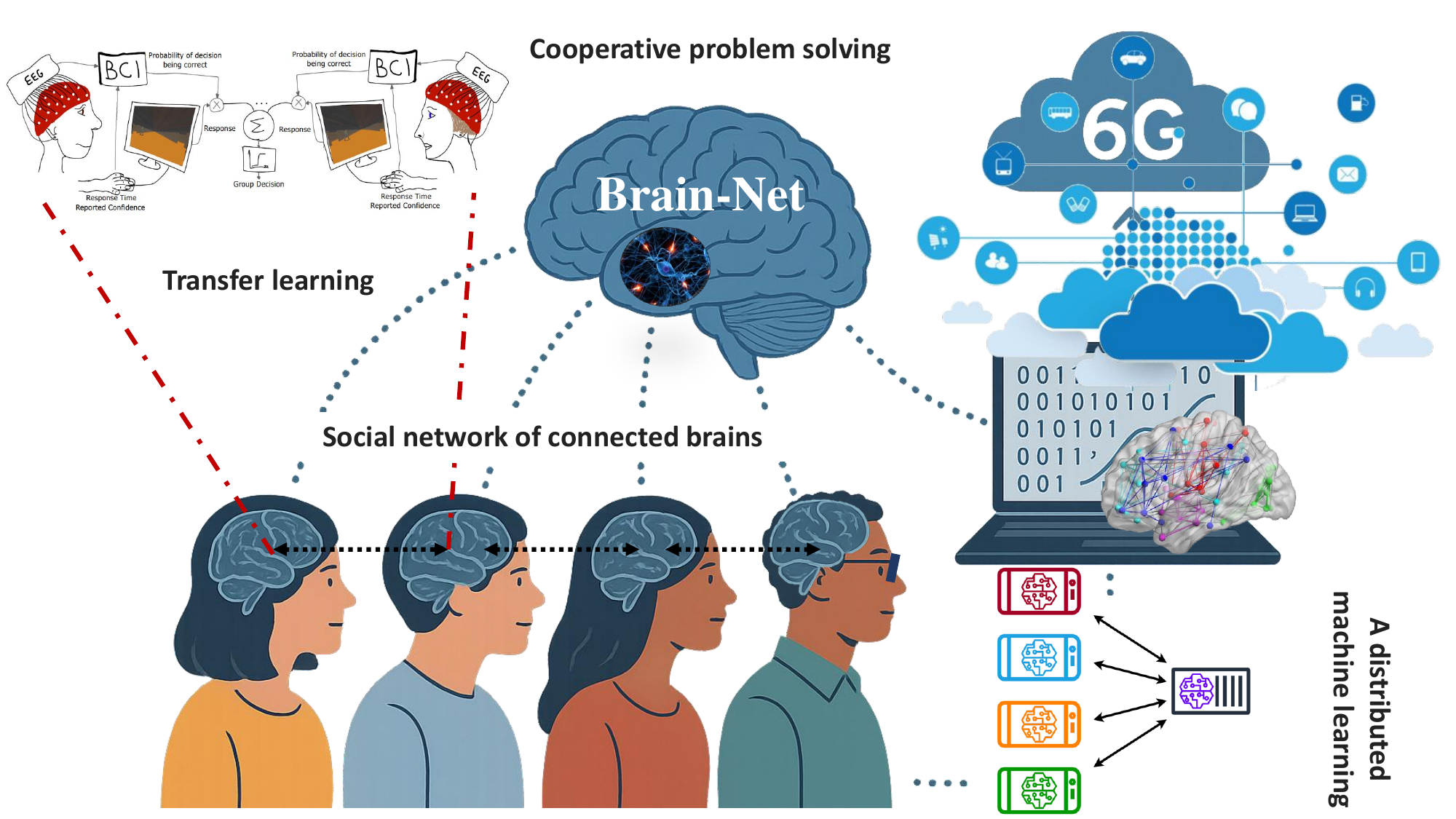}
  \caption{Brain-net and internet of minds powered by 6G and cloud intelligence.}
   \label{fig:Scenario2}
\end{figure}

Recently, researchers have introduced new concepts such as the Internet of Brains (IoB), Internet of Thoughts (IoTh), Internet of Thinking (IoTk), and Internet of Creation (IoC)~\cite{209}. The IoB envisions a network connecting human brains for direct communication and shared cognition. The IoTh extends this by allowing thoughts, emotions, and knowledge to be exchanged between humans and AI systems. The IoTk combines BCIs with real-time processing and AI to enhance thinking, decision-making, and the control of smart devices through neural signals. The IoC focuses on connecting human creativity with AI, enabling people to create and modify digital content using brain activity or enhanced cognitive abilities. 6G technologies will be the key enabler of these brain-based communication systems. With ultra-low latency, high-speed data transfer, distributed AI, and strong security, 6G will transform these ideas from theory into practical applications for collective intelligence, creativity, and human–AI collaboration.

\subsubsection{How 6G can help}
While 5G network marks a substantial advancement in wireless technology, it falls short in supporting the hyper-connectivity required for futuristic BCI applications. 6G networks will play a central part in enabling such future ideas to become reality such as IoB, IoTh, IoTk, and IoC. Such networks demand unprecedented levels of speed, reliability, latency, and convergence between biological and digital systems—functions 6G is uniquely designed to provide. This is how 6G can technically and accurately make such neurotechnological environments a reality \cite{saad2019vision}.

As demonstrated in Brain-Net experiments \cite{jiang2019brainnet}, real-time B2BI requires near-instantaneous transmission of neural data to support synchronous collaboration. 6G is expected to achieve latencies as low as 0.1 milliseconds—an order of magnitude lower than 5G ( $\sim $ 1 ms). This ultra-low latency is crucial for real-time decoding and encoding of neural signals, seamless communication between multiple participants in B2BI setups, and preventing time-lag induced misinterpretations in collaborative cognitive tasks.

BCI systems have high data rates and bandwidth for neural signal transmission \cite{dang2020should}. EEG and TMS-based systems generate complex, high-dimensional data. For effective B2BI and IoTh applications, this data must be transmitted without degradation or delay. 6G aims to provide terabit-per-second data rates, enabling high-fidelity transfer of raw or preprocessed neural signals, real-time brain state synchronization between individuals or with AI, and support for multiple simultaneous brain-based data streams.

IoTh and IoTk systems aim to merge human cognition with cloud and edge AI \cite{zhang20196g}. 6G’s architecture is designed to support distributed intelligence, leveraging AI at the network edge to process neural data close to the user for real-time feedback, enable adaptive BCIs that learn from user behavior and facilitate collaborative decision-making in multi-brain networks through context-aware AI agents. 6G will support massive machine-type communications, enabling synchronization of hundreds or thousands of interconnected brain interfaces and smart devices \cite{letaief2019roadmap}. This is essential for multi-user Brain-Nets enabling swarm intelligence, IoB systems in which entire populations can share sensory or cognitive data and IoC scenarios where distributed creative inputs are pooled in real time.

6G’s adoption of the THz spectrum (0.1–10 THz) will provide not only ultra-high bandwidth for neural data transfer but also potential for in-device bio-sensing. Terahertz (THz) waves can be used to noninvasively monitor brain activity or blood flow in real time. As well, hybrid THz communication and sensing could improve EEG/TMS integration accuracy. Brain data is deeply personal, and its transmission requires strong security guarantees \cite{nguyen20216g}. 6G is expected to integrate quantum communication protocols and blockchain-based decentralized security to ensure the protection of cognitive privacy, verification of neural input sources in Brain-Net systems, and the prevention of neurohacking in IoTh and IoB environments.

\subsection{BCI biometrics for security and privacy}
\subsubsection{Motivation}

Traditional authentication methods such as passwords, personal identification numbers (PINs), and pattern or gesture based locks—rely heavily on users’ memory ~\cite{lal2016review}. Passwords remain the dominant mechanism, yet managing multiple credentials across various accounts imposes a high cognitive load, especially for the elderly and cognitively impaired. These methods are also prone to dictionary, brute force, phishing, and social engineering attacks~\cite{babulal2023authentication}.

To address password fatigue, biometric methods like palm print and retina recognition have been proposed~\cite{yusuf2020survey}, but they are not universally accessible. Fingerprint recognition may fail for users with hand conditions, facial recognition is difficult for the visually impaired, and voice authentication is unsuitable for those with speech or hearing disorders. Moreover, biometric data can be forged—fingerprints or irises copied from photos and voices replayed for unauthorized access. Given these challenges, research has turned to Brain–Computer Interface (BCI)-based authentication, which leverages the uniqueness of individual brain activity patterns~\cite{finn2015functional}. Unlike conventional or physical biometrics, BCI authentication requires neither memorization nor physical input. Users simply verify their identity via neural signals captured by a BCI headset~\cite{kolivand2019brain}. This inclusive approach benefits people with disabilities and can be combined with other modalities, such as facial verification~\cite{9855859}, for multimodal security. Potential applications include financial services, aerospace, healthcare, and smart-home access control.

\subsubsection{How 6G can help}

The implementation of BCI biometric authentication over 5G networks poses several challenges, especially under extreme mobility conditions. For instance, authenticating to an online service using brainwave data at speeds beyond 500 km/h the upper limit supported by 5G demands advanced mobility management techniques (e.g., handover algorithms and mobile routing protocols) to maintain stable connectivity. By 2030, machine type devices are expected to reach 97 billion~\cite{bhat20216g}, potentially congesting 5G networks and causing latency or packet loss during authentication. Despite 5G’s advanced encryption features~\cite{ahmad2019security}, security vulnerabilities remain, particularly as the growing number of devices increases exposure to DDoS attacks and inter-slice threats within network slicing~\cite{de2023survey}. For example, an attacker could exploit a weaker slice to intercept BCI authentication data.

Emerging 6G technologies will overcome many of these limitations. With a higher frequency spectrum, 6G will support faster and more reliable brain data transmission, enhancing real-time BCI authentication. Beyond 5G techniques like beamforming and massive MIMO, 6G will integrate non-terrestrial satellite communications through Low Earth Orbit (LEO) constellations, offering global coverage with minimal latency~\cite{gustavsson2021implementation}. Moreover, 6G security will benefit from Post-Quantum Cryptography (PQC)~\cite{partala2021post}, which protects against quantum algorithms like Shor’s that can break RSA, ECC, and Diffie-Hellman~\cite{shor1999polynomial,kumar2022post}.

Additionally, combining 6G and blockchain enables decentralized BCI based authentication, mitigating single point failures and enhancing user privacy~\cite{10382044}. The ultra low latency of 6G will accelerate blockchain transactions, improving consensus and reliability. With full end-to-end automation via ZSM, 6G networks will also exhibit self healing and self-defending capabilities, autonomously detecting and mitigating threats such as intrusions or DoS attacks in real time.
In light of the limited research studies on brain digital twin and the complete absence of prior work in internet of minds, we summarize in~\tablename~\ref{tab:RE} the requirements of all BCI-enabled use cases and how 6G can support them.

\begin{table*}[!h]
\centering
\caption{Key requirements and 6G enhancements for each BCI-enabled use case}
\label{tab:RE}
\resizebox{\textwidth}{!}{%
\begin{tabular}{@{}lll@{}}
\toprule
\textbf{BCI Use Case} &
  \textbf{Requirements} &
  \textbf{6G Enhancements} \\ \midrule
Brain Digital Twin &
  \begin{tabular}[c]{@{}l@{}}- Multi-modal neural data for detailed and accurate emulation\\ of brain activity.\\ - High resolution and data rate for brain signal modeling.\\ - Real time synchronization between different modalities\\ (e.g. EEG with motion capture).\\ - High frequency synchronization between the physical brain \\ and its replica states.\end{tabular} &
  \begin{tabular}[c]{@{}l@{}}- THz bandwidth for high-resolution and real-time transmission of large, \\    multi-modal neural data\\ - URLLC ensures precise synchonization between the physical \\    and digital brains.\\ - IEC ensures real-time processing of high-dimensional brain data\\    at the edge.\end{tabular} \\ \midrule
Brain Immersion Systems &
  \begin{tabular}[c]{@{}l@{}}- Low-latency immersive interaction\\ - High data throughput for both BCI and XR data\\ - Tactile and brain feedback loop\\ - Reduce dependency on the cloud ande enhance privacy\end{tabular} &
  \begin{tabular}[c]{@{}l@{}}- Sub-ms latency between brain signals and XR environment events.\\ - Joint communication and haptics enhances user experience\\ - Adapts XR settings based on real-time cognititve and emotional state.\\ - AI-embedded edge processes brain and XR data simultaneously and locally\end{tabular} \\ \midrule
\begin{tabular}[c]{@{}l@{}}Brain Artificial Intelligence\\ Interface for Assistive Healthcare\end{tabular} &
  \begin{tabular}[c]{@{}l@{}}- Instant motor intent decoding and real time \\ control of devices for disabled people (wheelchairs, \\ robotic limbs, exoskeletons).\\ - Continuous and context-aware BCI operation across \\ dynamic and heteregenous environements.\end{tabular} &
  \begin{tabular}[c]{@{}l@{}}- Ultrafast wireless transmission of large brain signals\\ - Edge intelligence for real-time brain data processing.\\ - 6G-network slicing ensures dedicated, prioritized\\ and fault-tolerant service.\\ - 6G supports terahertz bands and higher mobility, enabling\\ reliable connectivity during user movement.\end{tabular} \\ \midrule
Brain-Net and Internet of Minds &
  \begin{tabular}[c]{@{}l@{}}- Multi-user, bidirectional and real-time \\ neural interfaces for cognitive collaboration\\ - Semantic neural signal interpretation\\ - Neuro-signal synchronization across brains\\ - Cogtintive context modeling\\ - Ultra-high uplink brandwidth\end{tabular} &
  \begin{tabular}[c]{@{}l@{}}- Massive mMTC allows seamless and synchronous\\   multi-brain interaction.\\ - High data rate transfer of dense neural signals\\ - Real-time decoding of brain signals and semantic thoughts locally\\ - Preserves identity and intent integrity during thought transmission\\ - Enable collaborative decisions in muli-brain networks\\ using context-aware AI agents.\end{tabular} \\ \midrule
BCI Biometric Authentication &
  \begin{tabular}[c]{@{}l@{}}- High-resolution and encrypted EEG signal capture\\ - Continuous identity verification\\ - Privacy-preserving edge analytic\end{tabular} &
  \begin{tabular}[c]{@{}l@{}}- URLLC and edge computing enables sub-second identity verification\\ - Ensure brain biometric templates remain secure long time \\ via post-quantum cryptography (PQC).\\ - Ensures cognitive-privacy via blockchain and advanced distributed\\ edge-based learning algorithms (FL, SL and SFL). \\ - 6G-based AI supports on device biometric authentication\end{tabular} \\ \bottomrule
\end{tabular}%
}
\end{table*}

 \section{Standardization Roadmap for 6G-enabled BCIs} 
\label{sec:Projects/Standards for 6G-enabled BCI}
To achieve a coherent convergence between BCI systems and 6G networks, two-layer standardization efforts are required: (1) BCI-focused standards that define device, signal, and data interoperability, (2) 6G-focused standards to establish the real-time, low-latency, and intelligent communication needed for next-generation cognitive and neural networks. ~\tablename~\ref{tab:standards} summarizes the main standardization efforts related to BCI and 6G convergence.

\begin{table*}[htbp]
\centering
\caption{Standardization Efforts Relevant to BCI and 6G Integration}
\label{tab:standards}
\small
\resizebox{\textwidth}{!}{
\begin{tabular}{|p{3cm}|p{3.5cm}|p{1.8cm}|p{4.2cm}|p{4.2cm}|}
\hline
\textbf{Organization / Body} & \textbf{Standard / Initiative} & \textbf{Status} & \textbf{Focus / Description} & \textbf{Relevance to BCI–6G Integration} \\
\hline
ISO/IEC JTC 1/SC 43 Brain–computer interfaces & ISO/IEC 8663 – Vocabulary for BCIs & Published (Sept 2025) & Specifies unified terminology and taxonomy for BCI systems. & Establishes a standardized vocabulary to ensure interoperability between BCI devices and communication systems. \\
\hline
ISO/IEC JTC 1/SC 43 & Hardware Interfaces and Protocols for BCIs & Proposal (2025) & Specifies data formats and interface protocols for neural signal exchange. & Facilitates data transmission between BCI sensors and 6G communication modules. \\
\hline
IEEE Brain & IEEE P2731 – Unified Terminology for Brain–Computer Interfaces & Ongoing & Defines a comprehensive BCI glossary, standard functional BCI model and consistent data structures. & Enables uniform representation of neural data across cloud/edge 6G infrastructures. \\
\hline
IEEE Brain & IEEE P2794 – Reporting of In Vivo Neural Interface Research & Approved (2022) & Standardizes how neural data and experimental setups are documented. & Improves reproducibility and cross-system compatibility for BCI data on 6G edge networks. \\
\hline
ITU-R (IMT-2030/6G Vision Framework) & Integrated Sensing and Communication (ISAC) & Approved (2024) & Incorporates sensing and communication in one unified 6G system. & Enables real-time neural sensing, localization, and feedback crucial for closed-loop BCI. \\
\hline
3GPP (Release 19 and Beyond) & 6G Advanced & Under Study & Extends URLLC, mMTC, and AI-native networking for 6G. & Supports massive, low-latency BCI data transfer and distributed cognitive communication. \\
\hline
China NMPA (Medical Device Standards) & YY/T 1987-2025: Brain-Computer Interface Medical Equipment-Terminology & Released (2025) & Define a unified terminology and framework for BCI medical devices to standardize research, development, regulation, and clinical applications. & Ensures reliability and compliance for 6G-connected BCI health applications. \\
\hline
\end{tabular}
}
\end{table*}

\section{Open Challenges and Future Directions} 
\label{sec:Open Challenges and Future Directions}
The convergence of BCI and 6G presents a promising technological advancement expected to shape the future of communication. However, several challenges must be addressed to ensure its effective implementation. In this section, we explore these challenges in detail.

\subsection{Limited 6G and BCI datasets} The heterogeneity of brain datasets is a primary issue in the development of generalizable and reliable brain signal decoding models. Present datasets are incredibly diverse as regards the number and spatial distribution of electrodes, the number of subjects involved, experimental paradigms, BCI modalities (e.g., P300 and/or SSVEP), and the basic neural measuring paradigms (e.g., EEG or ECoG). Furthermore, variations in recording protocols, experimental settings, and trial numbers complicate model generalization across studies. Domain shifts due to heterogeneity typically disrupt transferability and scalability of the decoding models. Therefore, to develop a 6G-enabled ML/DL BCI model, two key data types are required: (1) realistic 6G channel data (e.g., path loss, interference, bandwidth, etc.) and (2) brain data. However, since 6G is still in its early research phase, real-world 6G datasets are very scarce. The only realistic dataset publicly available in the literature is DeepSense6G~\cite{deepsense6g} from the Wireless Intelligence Lab. In response, researchers can use existing simulated 5G/6G datasets that provide insights into THz communication, massive MIMO, and ultra-reliable low-latency communication (URLLC). We can cite, for example, DeepVerse6G~\cite{deepverse6g} and DeepMIMO~\cite{deepmimo}. This ensures that BCI data transmission simulations reflect real-world 6G constraints. The second challenge is related to the BCI dataset. Many of BCI dataset are collected in research labs and medical institutions and are not publicly available due to privacy concerns. The existing dataset are very limited, sometimes small and noisy. The well-known BCI dataset in the literature is BCI Competition Datasets (I-IV)~\cite{BCICompetition}. Although these datasets are widely used and allow performance benchmarking, they include only 9 subjects and 2–5 sessions. PhysioNet EEG data~\cite{PhysioNet}, and Open BCI community datasets~\cite{OpenBCI} are two well-known sources for BCI-related datasets. Given these challenges, we encourage researchers and industry stakeholders to publicly share high-quality 6G and BCI datasets to support and advance future research in this emerging field.

\subsection{Wireless Communication Constraints} Terahertz band communication is a defining characteristic that sets 6G apart from other wireless technologies. Compared to existing networks, it has the potential to provide ultra reliable low latency and significantly higher data throughput. However, it also presents several challenges, including fading channels, interference, signal attenuation, and substantial path loss. In light of these issues, delays in transmitting brain signals can affect real-time control of brain-controlled devices. Moreover, interference could introduce errors that result in incorrect commands. Furthermore, high path loss forces BCI devices to consume more power, leading to energy efficiency concerns and reduced device usability. While, as discussed above, 6G will introduce cutting-edge solutions to reduce their impact, it cannot completely eliminate these constraints due to the fundamental laws of physics.

\subsection{Dataset Annotation Issues} Annotating brain data requires a multidisciplinary team, including neuroscientists, neurophysiologists, data scientists, signal processing specialists, and healthcare professionals. Performing this task manually is inherently difficult, time-consuming, and expensive, particularly for large scale and complex dataset. Moreover, in some scenarios, participants self-annotated the EEG signals (self-report), introducing a significant subjective component that can lead to inconsistent or inaccurate labels~\cite{cimtay2020investigating}. Furthermore, due to the uniqueness of the human brain, EEG signals can vary significantly from one individual to another. This variability can lead to differences in how experts interpret and label the same EEG data. As a result, experts can assign different labels to the same EEG segment, leading to inconsistencies in the labels assigned~\cite{li2019target}. 

Labeled data will still be useful in BCI; however, due the aforementioned challenges, the focus will shift toward techniques that can help BCI systems learn from brain signals without relying on explicit labels. In this regard, a promising technique is Self-Supervised Learning (SSL), which is beginning to gain more attention among researchers for its application in the EEG field~\cite{9837871}. With SSL, the BCI model can extract effective representation from unlabeled brain samples instead of directly training end to-end models through labeled brain sample. Likewise, Unsupervised Learning (UL)~\cite{huebner2018unsupervised}, is another approach that does not require labeled data at all and thus eliminates the need for annotation. It consists on extracting information form brain signals when a user performs mental tasks without predefined labels. The UL-driven BCI model identifies patterns, cluster similar signals and detect anomalies (unusual brain activities), enabling adaptive BCI control. Another technique to consider is Reinforcement Learning (RL)~\cite{girdler2022neural} where an adaptive controller, known as the agent, interacts with the human brain, which represents the environment, and continuously adjusts its behavior (agent) to enhance performance. The effectiveness of the BCI system is measured through cumulative rewards, which are assigned based on how well the BCI achieves its intended task, such as controlling a robotic arm or selecting commands through brain activity.

\subsection{Imbalanced and Non-IID Data}  
Naturally, participants in BCI trials do not generate an equal number of brain signal recordings for all cognitive states or motor imagery tasks, creating imbalanced data. Training a BCI model on such data can develop bias towards the predominant class, and fail to recognize rare classes. Moreover, inter-subject variability (each person’s brain activity is unique), inter-session variability (EEG signals can change across different recording days) and task variability (brain response to the same task can differ based on mental state, electrode placement or external distractions) make BCI data non-uniform~\cite{saha2017evidence}. 

Data augmentation is one of the widely used techniques to mitigate class imbalance in BCI. For example, in~\cite{krell2018data}, the authors suggest spatial augmentation using electrode position shifts and rotations. The authors in~\cite{um2017data}, used a random permutation of the signal segments to generate new EEG data. Recent studies have also explored GAN-based approaches for generating EEG signals that mimic real ones~\cite{fahimi2020generative, fan2020eeg}. Nevertheless, this requires large training sets and may introduce unrealistic noise or artifacts. Another approach is the resampling method, which involves removing samples from the majority class and increasing the minority samples until a balance is reached~\cite{varotto2021comparison}. However, this can lead to information loss or overfitting, and performance remains highly dependent on the classifier and metric choice. Therefore, selecting classifiers (e.g. ensemble learning) and performance metrics (e.g. area under the curve or balanced accuracy) that are less sensitive to class bias is crucial for reliable evaluation~\cite{mienye2022survey, varotto2021comparison}. While these techniques can improve data imbalance, they cannot fully address challenges related to variability, signal fidelity, and generalization, highlighting the need for adaptive, task-aware, and subject-specific augmentation strategies.

Beyond class imbalance, the non-IID nature of BCI data poses further difficulties. To address this, three main approaches can be applied: data-based, algorithm-based and system-based methods~\cite{zhu2021federated}. In data-based techniques, data sharing can handle non-IID data in distributed BCI systems, where the server maintains a subset of data that represents all possible BCI data types to train the global model. Then, each client (BCI device) receives a small, randomly selected subset to integrate with its local data during training. However, sharing data with clients compromises the privacy-preserving principles of distributed learning. In contrast, algorithm-centric approaches adapt the learning algorithm to enhance the BCI model robustness against non-IID variations. In this category, we find knowledge distillation in which a powerful global model is trained on diverse EEG data from multiple BCI users in a distributed manner. Then, each user receives compact student model that is distilled from the global model’s knowledge~\cite{hinton2015distilling}. The student model is then locally adapted using the user’s specific EEG data.

Finally, system-based methods focus on optimizing the learning architecture to handle non-IID data, for example, by clustering clients with similar data distributions for collaborative training. However, since EEG data is sensitive and private, clustering can not rely on raw data. Therefore, the main challenge is to determine an effective method to measure data similarity while preserving privacy. Existing studies have proposed (1) Loss-based similarity, where clients with similar loss values are grouped together~\cite{kopparapu2020fedfmc}, and (2) Model weights similarity, where clients with similar updates are assumed to have similar data distributions~\cite{briggs2020federated}. However, these techniques remain limited by noisy or varied updates that cannot accurately reflect true data similarity. Moreover, they introduce additional computation and communication overhead. Consequently, designing a reliable privacy-preserving similarity metrics that capture the intrinsic structure of EEG data without compromising performance or confidentiality is of paramount importance.

\subsection{Legal and Ethical Implications}

The clinical translation of neurotechnology, particularly BCIs, faces persistent challenges~\cite{vlek2012ethical}. Despite promising trials, a wide gap remains between experimental success and real-world adoption. A major limitation is the lack of rigorous randomized controlled trials (RCTs) to confirm safety and efficacy, complicating regulatory approval. Progress is further hindered by limited infrastructure, nonstandard protocols, and trial designs unsuited to neurotech applications. The growing market for unregulated direct-to-consumer (DTC) devices also raises safety and ethical concerns, especially when repurposed for medical use.

High costs, technical complexity, and reliance on MRI or EEG systems restrict accessibility, particularly in low-resource settings. Most neurotech data originates from high-income regions, resulting in biased models and reduced inclusivity. To expand access, researchers are developing portable, low-cost, and autonomous systems—such as wearable BCIs and closed-loop neuromodulation—to minimize clinical dependence.

Nonetheless, safety and ethical concerns persist. Invasive devices risk infection or tissue damage, while non-invasive systems may still cause discomfort or burns. Neural data introduce privacy, consent, and dual-use risks, including potential misuse for surveillance or manipulation. Global organizations such as the OECD, UNESCO, UNIDIR, and GESDA are addressing these issues by establishing ethical and regulatory frameworks. Integrating neuroethic education into research and clinical training, along with stronger public engagement, is essential to ensure responsible, transparent, and inclusive neurotechnology development.

\section{Conclusion}
\label{sec:Conclusion}
With the advancement of wireless communication technologies such as 6G, invasive and non-invasive BCIs can be made even more versatile and powerful. 6G-enabled BCIs have the potential to revolutionize a wide range of industries and improve the quality of life for people around the world. This paper is the first in the literature that provides a comprehensive overview on how the integration of these technologies could progress, examining what opportunities this convergence creates and how it can drive innovation. To this end, we present an in-depth exploration of the key 6G technical aspects for BCI integration, including intelligent physical layer, resource management, intelligent edge computing, privacy and security, as well as zero trust system management. Additionally, we review the practical and emerging use cases enabled by this convergence, such as brain digital twin, BCI biometric authentication, brain-net and internet of minds. We also outline potential frameworks and standards for consistent and interoperable implementation of 6G-enabled BCIs. Finally, we discuss some challenges and possible future directions that deserve further investigation. 
\bibliographystyle{IEEEtran} 
\bibliography{arXiv/References}

\end{document}